\documentclass[journal]{IEEEtran}

\IEEEoverridecommandlockouts
\usepackage{url}
\usepackage{footnote}
\usepackage{tablefootnote}
\usepackage{bm}
\usepackage{amsmath,amssymb,amsfonts}
\usepackage{diagbox}
\usepackage{csquotes}
\usepackage{mathtools}
\usepackage{algorithmic}
\usepackage{graphicx}
\usepackage{placeins} 
\usepackage[linesnumbered,ruled,vlined]{algorithm2e}
\usepackage{float}

\usepackage{textcomp}
\usepackage{xcolor}
\usepackage{mathrsfs}
\def\BibTeX{{\rm B\kern-.05em{\sc i\kern-.025em b}\kern-.08em
    T\kern-.1667em\lower.7ex\hbox{E}\kern-.125emX}}

\usepackage{makecell}
\usepackage{cuted}

\usepackage{hyperref}
\ifCLASSINFOpdf
\else
\fi
\begin{document}
%
\title{Deep learning-based Edge-aware pre and post-processing methods for JPEG compressed images}
%
%
%

\author{Dipti Mishra,~\IEEEmembership{Member,~IEEE,}
        Satish Kumar Singh,~\IEEEmembership{Senior Member,~IEEE,}
        and Rajat Kumar Singh,~\IEEEmembership{Senior Member,~IEEE}
        \vspace{-8mm}
\thanks{D. Mishra, R. K. Singh, are with the Department
of Electronics \& Communication Engineering, Indian Institute of Information Technology, Allahabad, Prayagraj, India  (e-mail: \{rse2017502, rajatsingh\}@iiita.ac.in)
}
\thanks{ S. K. Singh is with the Department of Information Technology, Indian Institute of Information Technology, Allahabad, Prayagraj (sk.singh@iiita.ac.in)}

}

\markboth{THIS WORK HAS BEEN SUBMITTED TO IEEE TRANSACTIONS FOR CONSIDERATION}%
{Shell \MakeLowercase{\textit{et al.}}: Edge-Aware Image Compression using Deep Learning-based Super-resolution Network}
%



\maketitle

\begin{abstract}
  We propose a learning-based compression scheme that envelopes a standard codec between pre and post-processing deep CNNs. Specifically, we demonstrate improvements over prior approaches utilizing a compression-decompression network by introducing: (a) an edge-aware loss function to prevent blurring that is commonly occurred in prior works \& (b) a super-resolution convolutional neural network (CNN) for post-processing along with a corresponding pre-processing network for improved rate-distortion performance in the low rate regime.
  The algorithm is assessed on a variety of datasets varying from low to high resolution namely Set 5, Set 7, Classic 5, Set 14, Live 1, Kodak, General 100, CLIC 2019. 
  When compared to JPEG, JPEG2000, BPG, and recent CNN approach, the proposed algorithm contributes significant improvement in PSNR with an approximate gain of 20.75\%, 8.47\%, 3.22\%, 3.23\% and 24.59\%, 14.46\%, 10.14\%, 8.57\% at low and high bit-rates respectively. Similarly, this improvement in MS-SSIM is approximately 71.43\%, 50\%, 36.36\%, 23.08\%, 64.70\% and 64.47\%, 61.29\%, 47.06\%, 51.52\%, 16.28\% at low and high bit-rates respectively. With CLIC 2019 dataset, PSNR is found to be superior with approximately 16.67\%, 10.53\%, 6.78\%, and 24.62\%, 17.39\%, 14.08\% at low and high bit-rates respectively, over JPEG2000, BPG, and recent CNN approach. Similarly, the MS-SSIM is found to be superior with approximately 72\%, 45.45\%, 39.13\%, 18.52\%, and 71.43\%, 50\%, 41.18\%, 17.07\% at low and high bit-rates respectively, compared to the same approaches. A similar type of improvement is achieved with other datasets also.
\end{abstract}


\section{Introduction} \label{sec:Introduction}
\IEEEPARstart{E}{very} day an enormous amount of data is continuously transmitted and stored. Mostly the transmission happens through the web, and the images on the internet are of high resolution and larger in size. So to improve the efficacy of the transmission and storage, reducing the size of these files is a compulsion and an issue of big concern. Hence, image compression is widely used for storage \& transmission which aims at reducing the different types of redundancies present in the image. A higher compression ratio reduces the size of the image to be stored as bits-per-pixel (bpp) allotted decreases, but it heavily affects the reconstruction quality of the image due to the loss that occurred during compression. So, there is a trade-off between the compression ratio and the quality of the image. Among alternative compression formats, the Joint Photographic Experts Group (JPEG)~\cite{standard:jpeg94} standard is by-far dominant; currently, almost all pictorial images are stored and communicated in JPEG format. Given the extensive utilization of JPEG, modern SmartPhone \& tablet devices commonly include hardware support for JPEG compression \& decompression. Methods that can improve compression performance, while maintaining compatibility with the JPEG format, are, therefore, of strong research interest. 
Theoretically, JPEG, the standard algorithm, which is a discrete cosine transform (DCT) based emphasizes low-frequency components while processing the image for the task of image compression. 
JPEG2000~\cite{jpeg2000}, on the other hand, being lossless \& lossy both, is a region of interest (ROI) based compression scheme based on multi-scale decomposition into sub-bands through discrete wavelet transform (DWT). 
As compared to JPEG, JPEG2000 produces only ringing effects near the edges in the image. However, at high bit-rates, these artifacts become less visible or almost imperceptible.
Later on, many techniques like BPG\footnote{\url{https://bellard.org/bpg/}} \& WebP\footnote{\url{https://developers.google.com/speed/webp/}} came into existence which provided a significant quantitative \& qualitative improvement with respect to reconstructed image quality.
Nowadays, deep learning is producing interesting \& eye-catching results in the field of biometrics \& computer vision, it has started proving itself in the image compression domain also. Some of the best compression algorithms are discussed in the next sections.

\subsection{Deblocking and Post-processing methods}
Various post-processing modules have been proposed in the literature. Modeling of the image-prior, based on maximum a posteriori (MAP) criteria is used in~\cite{sun}, but it is found to be quite more expensive than handcrafted models. The post-processing method in~\cite{foi} based on the adaptive DCT method is quite good, but failed to generate sharp images. The notion of Wiener filtering for denoising used in~\cite{bm3d} was novel, however, the images generated contained highly visible noise at the edge regions. Similarly, the modeling of image prior and quantization is found to be expensive in~\cite{overlap}. Ren et al.~\cite{ren} combined the local sparsity image prior model and non-local similarity model for artifacts reduction using low-rank minimization and patch clustering proposed in~\cite{sun} and~\cite{overlap} respectively. Dictionary learning and variational regularization used in~\cite{dictv} for restoring images outperformed the other decompression method but consumed substantial computational time. Weighted nuclear norm minimization (WNNM) concept used in~\cite{wnnm} being complex, outperformed Dabov et al.~\cite{bm3d} approach while preserving the local structure and reducing visual artifacts taking more computation time.
The deblocking method, given in~\cite{concolor} using a non-convex low rank constrained model was a good optimization of the problem. The denoising method proposed by Zhang et al.~\cite{dcnn} helped in reducing artifacts but with the lack of retaining edge information in reconstructed images. The benchmark deblocking method ARCNN, proposed by Dong et al.~\cite{arcnn} outperformed the above-discussed methods used for image compression. 
All these deblocking and post-processing models aimed to improve the reconstruction quality while ignoring the joint optimization of the traditional codec with encoder-decoder structure.

\subsection{Down and Upsampling based CNN methods}
Recent work~\cite{framework,virtual} has demonstrated promising results with deep learning methodologies for compression. Specifically, the work in~\cite{framework} proposed an \enquote{end-to-end} compression framework wherein pre \& post-processing DNN is introduced around a standard compression codec. However, the training procedure did not take care of high-frequency components (i.e. edges) due to optimization with mean square error (MSE) resulting in smooth reconstructed images.
Moreover, the author did not present the rate-distortion analysis for the JPEG codec, which is needed to compare the compression performance based on bit-rates.
Like~\cite{framework}, the approach in~\cite{virtual} uses jointly trained CNNs for the pre-processing and post-processing stages. These stages incorporate resolution reduction only for low JPEG quality factor (QF) and maintain resolution for higher QF. 
On the other hand, the work in~\cite{feng} is based on investigating different downsampling networks and super-resolving or upsampling networks along with the different training strategies. The sequential training of downsampling CNN with upsampling CNN only has been considered, without the consideration of any codec in-between.

\subsection{Recent State-of-the-Art Algorithms}
In 2016, Toderici et al.~\cite{fullresolution} introduced an RNN based full resolution compression exploiting the principle of progressive encoding-decoding of images.
In 2017, Balle et al.~\cite{optimized} proposed to use generalized divisive normalization based local gain control method with the use of additive uniform noise to make non-differentiable quantization function a continuous function.
Theis et al.~\cite{theis} developed an algorithm to use a single autoencoder for providing variable bit-rate image compression by just tweaking with the size of latent space representation before quantization. Basically, the work in~\cite{fullresolution,rippel,johnston} is based on the calculation of the probability of binary representation for context-adaptive coding. Basically, Johnston et al.~\cite{johnston} extracted binary information for maximizing the entropy. 
In 2018, Li et al.~\cite{learning} came up with a compression model based on content prediction with the help of importance map generation through CNN.
Later on, Balle et al. extended the work in~\cite{optimized} by exploiting spatial dependencies for variable rate image compression in~\cite{balle2}.
Similarly, Lee et al.~\cite{lee} extended the work in~\cite{balle2} by proposing a context-adaptive entropy model by incorporating bit-consuming and bit-free contexts. 
On the other hand, the approach reported by Minnen et al.~\cite{minnen} was based on weighted auto-regressive and hierarchical priors for contextual image compression. Mentzer et al.~\cite{mentzer} proposed to focus on rate-distortion trade-offs by learning a conditional probability model of latent space distribution in a 3D-CNN autoencoder.
Later on in 2019\textquotesingle s, Choi et al.~\cite{choi} proposed a variable bit-rate image compression in the hierarchy of having one compression module for providing multiple bit-rates.
In early 2020\textquotesingle s, the model proposed by Yang et al.~\cite{yang} was based on multi-layer feature modulation which reduced the problem of lower performance at low bit-rates. Chen et al.~\cite{chen} proposed to learn a single CNN model at high bit-rates and then scaling the bottleneck by some factor for low bit-rates.
\begin{figure*}
\centerline{\includegraphics[width=\linewidth,height=8cm]{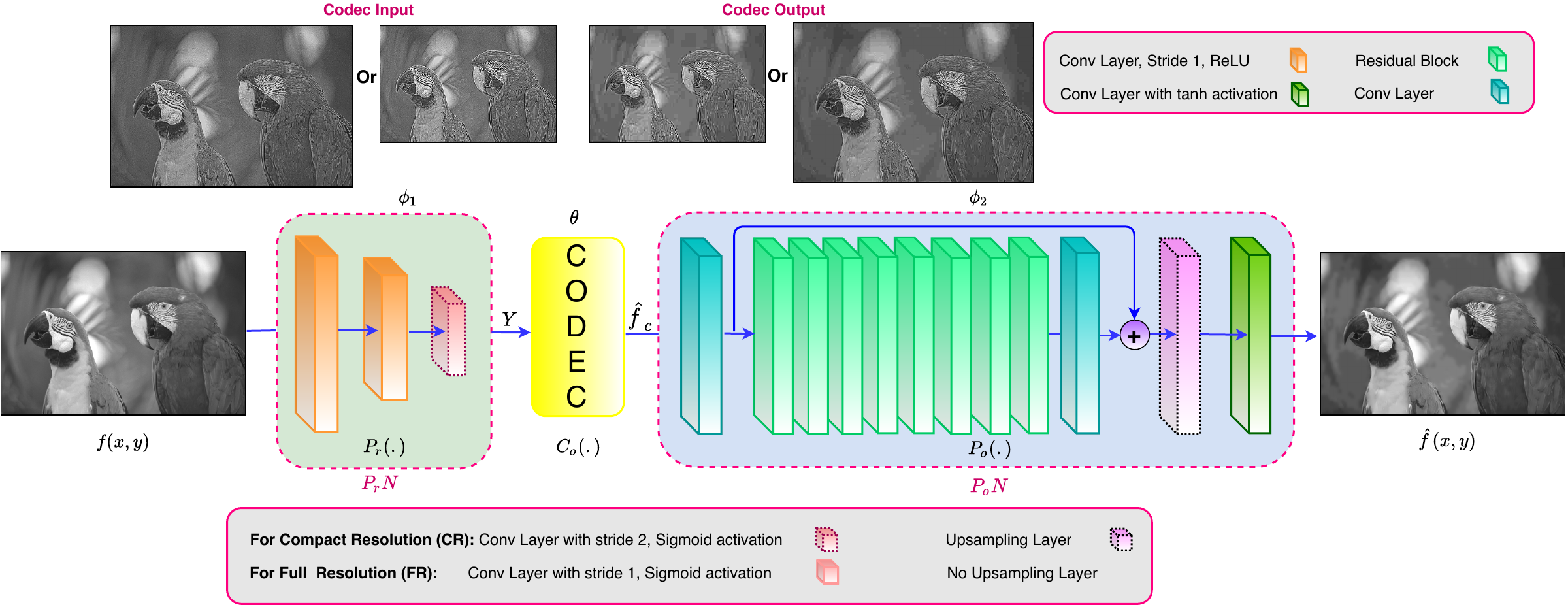}}
\caption{Detailed architecture of the proposed compression-decompression network.}
\label{fig:edsr21}
\end{figure*}

Motivated by the discussions on downsampling/upsampling based CNN methods, we utilize an \enquote{end-to-end} deep compression framework compatible with the traditional codec by utilizing pre \& post-processing DNNs. Different from prior works, however, we:
\begin{itemize}
    \item introduce an edge-aware loss function during compression, which significantly improves compression performance. The edge-aware loss function is motivated by the need to minimize edge distortion in compression, which the MSE loss function in~\cite{framework} does not emphasize.
    \item for the low bit-rate regime, utilize a super-resolution network in the post-processing stage.  
    \item prefer to use direct image-to-image learning over patch-to-patch learning criteria which is generally practiced in previous compression algorithms. The later scheme degrades the training performance, prediction efficiency, and lead to distortions in the output image.
   \end{itemize}
In the proposed approach, we utilize the enhanced deep super-resolution (EDSR) network~\cite{edsr}, which is a multi-scale network designed for specific super-resolution scaling that reduces model size and training time using residual learning compared with alternative approaches. The residual learning used in the super-resolution network helps in preserving the detail components.
Because the quantization in the codec represents a non-differentiable operation, we use progressive training that co-optimizes the pre \& post-processing modules in an \enquote{end-to-end} manner by appropriately approximating the codec. 
The super-resolution is chosen to improve the quality of the JPEG-compressed output by reducing blocking artifacts at the time of post-processing. In-fact, super-resolution which is a process of converting a low-resolution image into its high-resolution counterparts is exactly equivalent to image decompression which aims to improve the quality of the decoded image by adding more high-frequency details. Hence, the image super-resolution mechanism can be called a special case of an image compression-decompression algorithm.
Different from~\cite{framework,virtual}, the proposed network uses an edge-aware loss function to optimize the pre-processing module. In addition to that, it uses EDSR module instead of VDSR~\cite{vdsr} module (used in~\cite{framework}) \& 5 layer encoder-decoder network (used in~\cite{virtual}) as post-processing module. The proposed algorithm is quite different from the work reported in~\cite{virtual,feng} in terms of the pre-processing module and post-processing module exploited.
Moreover, the algorithm presented in~\cite{virtual,feng} exploited switching between one network to another when investigating compression performance from low bit-rates to high bit-rates, which does not seem optimal. Also, the author did not compare the work reported in~\cite{feng} with any DNN based image compression algorithms.

The remaining sections of the paper are organized as follows.
Section~\ref{sec:EdgeAwareDeepJPEG} presents the proposed approach covering details of the post-processing DNN \& the edge-aware loss function~\cite{edge2,edge} that we introduce. Section~\ref{sec:ExpDetails} discusses for the datasets used and other experimental setup details. In Section~\ref{sec:ablation}, comparison between different experiment strategies and modules has been shown.
Section~\ref{sec:comparison} throws light on the comparison of the final proposed algorithm with the state-of-the-art algorithms. Section~\ref{sec:conclusion} finally summarizes the effectiveness of the proposed algorithm reported in this paper.


\section{Edge-Aware Codec Compression} \label{sec:EdgeAwareDeepJPEG}
The proposed architecture with complete layer-wise details for the three separate modules is shown in Fig.~\ref{fig:edsr21}. The image to be compressed, $\mathbf{f}(x,y)$, goes through an image pre-processing network ($P_{r}N$) that generates the input image  $\mathbf{Y}=P_{r}(\mathbf{f},\bm{\phi_{1}})$ for the standard codec (here, JPEG), where $\bm{\phi_{1}}$ are the network weights, i.e, the learned parameters, of $P_{r}N$. Decompression is performed by first decoding the image from its coded JPEG representation to obtain the image $\hat{\mathbf{f}}_{c}=C_{o}(\mathbf{Y},\theta)$, where, $Co(.)$ represents the codec function, $\theta$ denotes the JPEG QF. The reconstructed image $\hat{\mathbf{f}}(x,y)$ is then obtained by post-processing the image $\hat{\mathbf{f}}_{c}$ through the post-processing network ($P_{o}N$); this operation is represented as 
$\hat{\mathbf{f}}(x,y) = P_{o}(\hat{\mathbf{f}}_{c},\bm{\phi_{2}})$, where $\bm{\phi_{2}}$ are the network weights of $P_{o}N$. In this process: (a) for $P_{E}~(CR)$ (proposed edge-aware compression algorithm based on compact resolution (CR) network), the image $\mathbf{Y}$ produced by $P_{r}N$ is a CR (scaled down in size by $2$ along each dimension by using a stride of $2$ in the final convolutional layer), \& $P_{o}N$ includes an upsampling layer that scales up the resolution by a factor of $2$ along each dimension \& (b) for $P_{E}~(FR)$ (proposed edge-aware compression algorithm based on full resolution (FR) network), the resolution is held constant by the processing by $P_{r}N$ \& $P_{o}N$. $P_{r}N$ uses an arrangement of layers identical to~\cite{framework}, except that the second convolutional layer consists of 32 filters instead of 64 filters in~\cite{framework}. Accordingly, to make the downsampled image to be as informative as possible, $P_{r}N$ uses stride 2 in the last CNN layer (different from Li\textquotesingle s). We defer using the batch normalization (BN) (as used in~\cite{framework}) for better generalizability, stable training, and consistent performance. Firstly, normalizing features with the mean and variance of batch during training generate artifacts, if the statistical characteristics of the training and testing datasets largely differ~\cite{edsr,kim}. Secondly, in PSNR oriented tasks, like image enhancement, image super-resolution \& image compression, BN gives rise to unpleasing artifacts when the CNN network is deeper. Thirdly, feature normalization reduces range flexibility which is also experimentally proven in~\cite{edsr}. Lastly, BN layers consume the same amount of memory as the preceding convolutional layers, this elimination also significantly reduces the network parameters \& GPU memory usage. The post-processing network $P_{o}N$ uses an adaptation of the EDSR network~\cite{edsr}, with an up-sampling layer at the output when a CR representation is used. The use of this upsampling layer is dropped when the FR representation is used. The upsampling layer in CR representation replaces the bicubic interpolation that was used in~\cite{framework} to recover a full-size image from CR representation. Fully connected layers have not been used in the proposed network making it a fully convolutional network (FCN) in nature.

\setlength{\parskip}{-2pt}
We stress upon the fact that, while the approach of reducing the codec input to a lower (compact) resolution is beneficial at low rates (correspondingly, lower reconstruction quality), at higher rates, better rate-distortion performance is obtained by retaining the original or full resolution instead of using the reduced resolution.
\setlength{\parskip}{-2pt}

Structural and textural details are the important features that are highly correlated to human visual perception. Training a DNN with MSE produces smooth images at the output while ignoring the fine structural features, which are also important details for human visual perception. The edges present in the images which are mainly responsible for structural features should also be considered and given some weight while training the network. Accordingly, the edge-aware loss function is defined as,
\begin{equation}
    \bm{\mathscr{L}}_{edge-aware}= \alpha \times \bm{\mathscr{L}}_{MSE} + \gamma \times \bm{\mathscr{L}}_{edge~weighted~MSE} \label{eq:edge}
\end{equation}
where, $\gamma=1-\alpha$ \& $\alpha$ decides the weight for the two components in Eq.~\eqref{eq:edge}. The first component in Eq.~\eqref{eq:edge} is basically MSE and the second component is the edge-preserving loss function (weighted MSE) which helps to restore edges or structural information in the image. Learning high-quality edges ensures the good quality of the reconstructed images. 

Accordingly, for the $P_{r}N$ module, a novel edge aware loss function for an image $\mathbf{f}$ is introduced as in Eq.~\eqref{eq:lr},
\begin{equation}
\begin{split}
      \bm{\mathscr{L}_{r}}(\bm{\phi_{1}},\mathbf{f})=\alpha \left \|P_{o}(\hat{\bm{\phi_{2}}},P_{r}(\bm{\phi_{1}},\mathbf{f}))-\mathbf{f}  \right \|_{2}^2 \\+   \gamma  \left \| \mathbf{E} \circ \left ( P_{o}(\hat{\bm{\phi_{2}}},P_{r}(\bm{\phi_{1}},\mathbf{f}))-\mathbf{f}  \right ) \right \|_{2}^2
      \label{eq:lr}
\end{split}
\end{equation}
where $\left \|  \cdot \right \|_2$ denotes the $\ell$-2 vector norm, $\mathbf{E}$ denotes the edge weight map (binary map representing edge at 1 \& no edge at 0) obtained by applying the edge detector to the input image, \& $\circ$ denotes pixel-wise multiplication.

The post-processing network $P_{o}N$ uses the MSE loss function as given in Eq.~\eqref{eq:lo} as,
\begin{equation}
     \bm{\mathscr{L}_{o}}(\bm{\phi_{2}},\mathbf{f})= \left (\left \|P_{o}({\bm{\phi_{2}}},C_{o}(P_{r}(\hat{\bm{\phi_{1}}},\mathbf{f}),\theta))-\mathbf{f} \right \|_{2}^2  \right )
     \label{eq:lo}
\end{equation}
\begin{figure}
\centerline{\includegraphics[width=8cm,height=3.5cm]{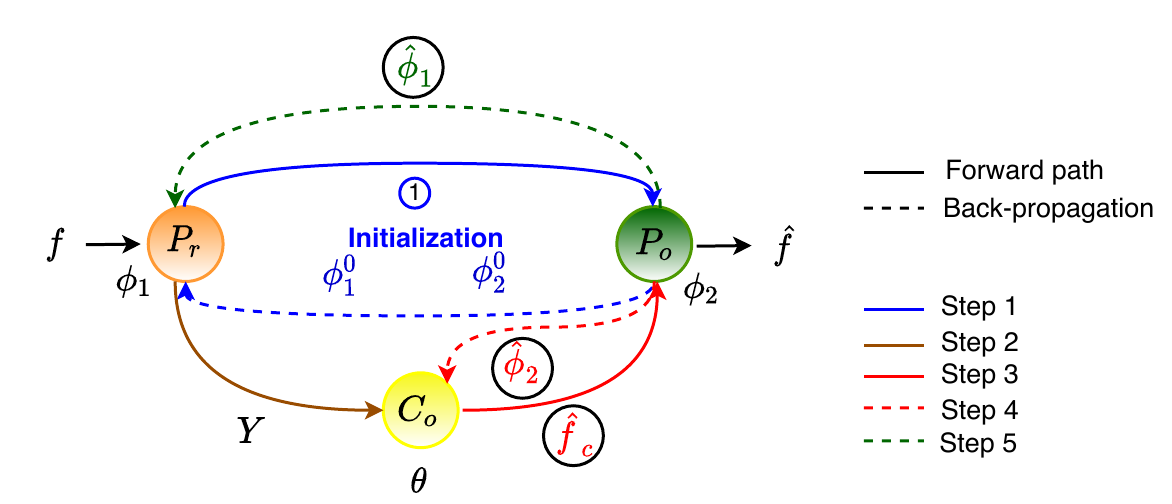}}
\caption{Learning flow for proposed algorithm.}
\label{fig:workflow}
\end{figure}
Also, Fig.~\ref{fig:workflow} clearly shows the progressive learning workflow for the proposed algorithm.
Initially, both the CNN modules are initialized with the learning parameters $\bm{\phi_{1}}^{0}$ \& $\bm{\phi_{2}}^{0}$ with training as a simple autoencoder, as shown in step 1. Then, the CR (in case of $P_{E}~(CR)$) or FR representation (in case of $P_{E}~(FR)$) is predicted using weights of $P_{r}N$ module as in step 2. This representation is compressed with JPEG for a particular QF, as in step 3.
Now, we define any arbitrary variable $\hat{\textbf{\textit{f}}}_{c}$, codec output, acting as input to $P_{o}N$ module. Accordingly, for an image $\textbf{f}$, codec output $\hat{\textbf{f}}_{c}$ can be written as in Eq.~\eqref{eq:upfc} as,
\begin{equation}
   \hat{\mathbf{f}_{c}}=C_{o}(P_{r}(\hat{\bm{\phi_{1}}},\mathbf{f}),\theta)   \label{eq:upfc}
\end{equation}
Hence, further to optimize $P_{o}N$ module for batch of images, the learning parameter $\bm{\phi_{2}}$ can be obtained as in Eq.~\eqref{eq:phi22},
\begin{equation}
  \hat{\bm{\phi_{2}}}=\arg \min_{\bm{\phi_{2}}} \left (\sum_{k=1}^{N} \left \| P_{o}({\bm{\phi_{2}}},C_{o}(P_{r}(\hat{\bm{\phi_{1}}},\mathbf{f}_{k}),\theta))-\mathbf{f}_{k} \right \|_{2}^2 \right ) \label{eq:phi22}
\end{equation}

For batch of images, Eq.~\eqref{eq:phi22} can be modified using using Eq.~\eqref{eq:upfc} to,
\begin{equation}
   \hat{\bm{\phi_{2}}}=\arg \min_{\bm{\phi_{2}}}  \left (\sum_{k=1}^{N} \left \| P_{o}({\bm{\phi_{2}}},\hat{\mathbf{f}}_{c_{k}})-\mathbf{f}_{k} \right \|_{2}^2 \right ) \label{eq:upphi2}
\end{equation}
where Eq.~\eqref{eq:upphi2}, implies that the codec output $\hat{\mathbf{f}}_{c_{k}}$ in Eq.~\eqref{eq:upfc} and ground truth $\mathbf{f}_{k}$ are then used to optimize the $P_{o}N$ training parameter $\bm{\phi_{2}}$ as in step 4. With $\bm{\hat{\phi}_{2}}$, $P_{r}N$ module is optimized by back-propagating error through $P_{r}N$ module to obtain $\bm{\hat{\phi}_{1}}$ in step 5 as in Eq.~\eqref{eq:compphi1} as,

\begin{equation}
\begin{split}
\hat{\bm{\phi_{1}}}=  \arg \min_{\bm{\phi_{1}}} \left( \alpha \sum_{k=1}^{N} \left  \| A  \right \|_{2}^2 \right )       +    \arg \min_{\bm{\phi_{1}}} \left ( \gamma   \sum_{k=1}^{N}\left \|   \mathbf{E} \circ A \right \|_{2}^2 \right )
\label{eq:compphi1}
\end{split}
\end{equation}
where, $A$ in Eq.~\eqref{eq:compphi1} can be written as, 
\begin{equation}
    A=P_{o}(\hat{\bm{\phi_{2}}},C_{o}(P_{r}(\bm{\phi_{1}},\mathbf{f}_{k}),\theta))-\mathbf{f}_{k}
\end{equation}
Since to optimize the weights of $P_{r}N$, codec is excluded in the back-propagation path, accordingly, on putting the value of $A$, finally it can be deduced that, 
\begin{equation}
\begin{split}
 \hat{\bm{\phi_{1}}}=\arg \min_{\bm{\phi_{1}}} \left(\alpha \sum_{k=1}^{N}\left \| P_{o}(\hat{\bm{\phi_{2}}},P_{r}(\bm{\phi_{1}},\mathbf{f}_{k}))-\mathbf{f}_{k}  \right \|_{2}^2 \right ) \\       +  \arg \min_{\bm{\phi_{1}}} \left (  \gamma \sum_{k=1}^{N} \left \|  \mathbf{E} \circ \left (P_{o}(\hat{\bm{\phi_{2}}},P_{r}(\bm{\phi_{1}},\mathbf{f}_{k}))- \mathbf{f}_{k} \right ) \right \|_{2}^2 \right ) 
 \label{eq:upphi1}
 \end{split}
\end{equation}
Note that codec is used in the forward path only to predict the input to $P_{o}N$. The flow can also be called predicting (testing) while training. In every epoch, the weights of $P_{o}N$ are fine-tuned according to the JPEG decoded output at a particular QF. Then the weights of $P_{r}N$ are fine-tuned by fixing the weights of $P_{o}N$.
Hence, it is clear that the training of both the CNN modules depends upon JPEG QF $\theta$, as all the intermediate outputs $\mathbf{Y}$ \& $\hat{\mathbf{f}}_{c}$ depends upon $\theta$. Hence, it is deduced that the codec JPEG also plays a role in CNN weight learning of $P_{r}N$ \& $P_{o}N$. Here, we have used the JPEG codec, however, the framework is generalized to use any codec. 

\section{Implementation Details} \label{sec:ExpDetails}
\subsection{Dataset}
The proposed compression-decompression model is trained on 400 images ($256\times256$) of the ImageNet dataset~\cite{imagenet}. In order to compare the results with~\cite{virtual}, we have assessed the algorithm on various datasets namely Live 1~\cite{live}, Set 14~\cite{set14}, Set 5~\cite{set5} \& Set 7 (same as built up in~\cite{virtual}). These datasets are having different attributes which also help in the better generalization of the proposed scheme. 
Live 1 dataset has twenty-nine diversified generic  PNG images with sizes ranging from $438\times634$ to $720\times480$. These images contain high resolution and high-quality, sharp images. Set 14 contains fourteen low-frequency PNG images with size ranging from $250\times361$ to $768\times 512$. Set 5 contains five edge centered luminance images. These are images with high-frequency domination where size varies from $280\times280$ to $512\times512$.
Apart from these three standard test sets, a test of 7 images named Set 7 is prepared (as in~\cite{virtual}) to compare the results with Zhao\textquotesingle s approach~\cite{virtual}. This special set contains both smooth and sharp images with size $256\times256$, $512\times512$ \& $512\times768$, and reported as challenging for image compression application. To compare with the recent techniques, the proposed algorithm is also assessed on Kodak benchmark dataset~\cite{kodak} and CLIC challenging dataset~\cite{clic}. 
Kodak dataset contains uncompressed true-color PNG images, with a size of $768\times512$ or $512\times768$. 
CLIC 2019 consists of high-resolution and high-quality images with size ranging from $512\times768$ to $2048\times1646$.
Moreover, to compare the deblockiness effect of the proposed algorithm, we have also tested Classic 5~\cite{classictest} \& General 100~\cite{general100} datasets. Classic 5 contains five BMP gray images of $512\times512$ size. General 100 contains a hundred images in BMP format with zero compression, which makes it highly suitable for the compression task. It contains good quality images with clear sharp edges but fewer smooth regions. The image dimensions range from $131\times112$ to $710\times704$. 

\subsection{Hyperparameter Tuning}
Both the CNN modules are trained for five iterations in each epoch. Adam optimizer is used to optimize the weights of the network that has shown very good performance as reported in~\cite{adam}. Initially, $\alpha=0.01$ was selected as the learning rate, but the loss observed was very high. Then we decreased the learning rate to $\alpha=0.001$, due to which loss also decreased substantially. On further decreasing the learning rate, we did not get much improvement. Hence, we finalized the learning rate at $\alpha=0.001$. 
The momentum parameters are selected as $\beta_1=0.9$, $\beta_2=0.999$ \& $\epsilon=1\times10^{-7}$, which are reported to yield good results as in~\cite{framework,edsr,dipti} to find the global minima.
The batch size depends upon the computational platform used for the implementation of the CNN based algorithm. It should fit into the memory specification of CPU or GPU based architecture. Generally, training on small batch sizes speeds up the learning process at the cost of low accuracy on account of noise in the training process. On the other hand, with large batch size, the learning process converges slowly with better prediction or estimation.
Accordingly, we have set a mini-batch size of 10, suitable for the used platform. We have finalized training epochs to be 50, as, after that, the performance was not improving. All the experiments have been conducted using Python with Tensorflow \& Keras deep learning framework on TITAN X (Pascal)/PCIe/SSE2 16 GB CPU RAM, 12 GB GPU enabled system.

\begin{figure}
\centering
{\includegraphics[width=0.48\textwidth,height=2.8cm]{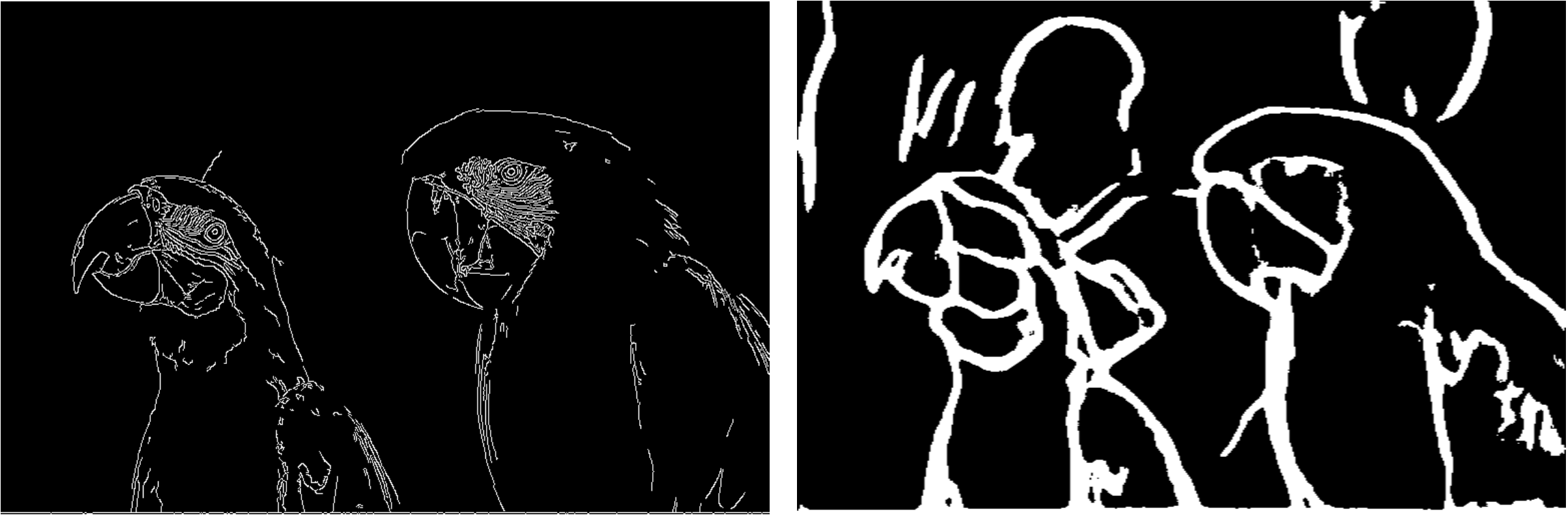}} 
\caption{Comparison of the performance of classical CED and learning-based HED on \enquote{Kodim 23} image of Kodak dataset. Clearly, HED produces strong edges and eliminate false edges which are not contextually required.} 
\label{fig:hed}
\vspace{-5mm}
\end{figure}

\begin{figure*}
\centering


\includegraphics[width=\linewidth,height=0.65\textwidth]{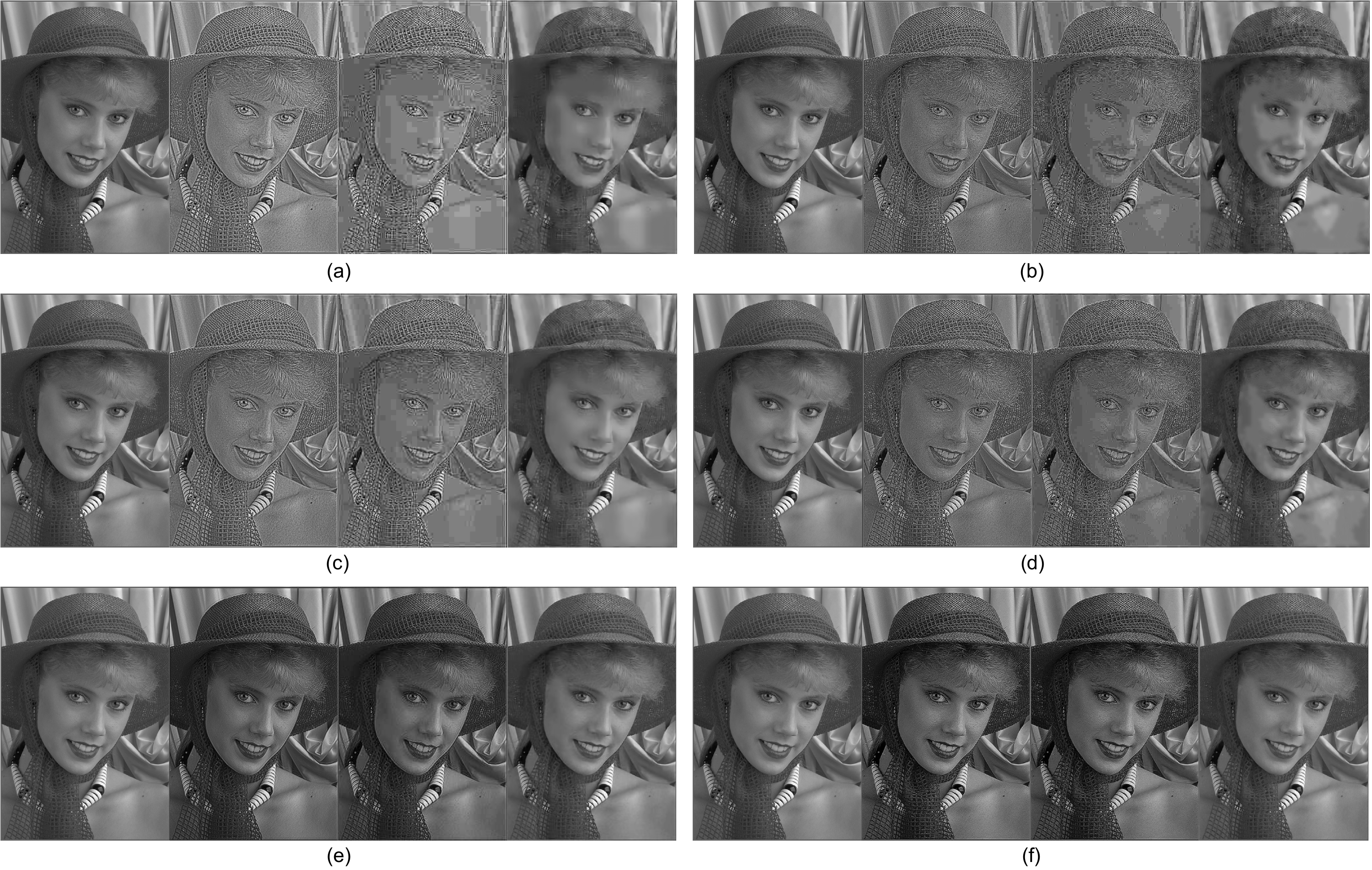}
\caption{(a), (c) \& (e): $P_{E}~(CR)$; (b), (d) \& (f): $P_{E}~(FR)$ compression performance on \enquote{Kodim 04} image of Kodak dataset at QF of 6 (row 1), 10 (row 2) \& 50 (row 3). Compression performance is (a) 6, 0.042, 27.19, 0.7644, 0.8789, 27.19, 0.1832; (b) 6, 0.1238, 28.79, 0.8466, 0.9116, 28.79, 0.3756; (c) 10, 0.0528, 28.46, 0.7931, 0.9135, 28.46, 0.2071; (d) 10, 0.1832, 32.01, 0.8954, 0.9525, 32.01, 0.4856; (e) 50, 0.1144, 31.71, 0.8554, 0.9656, 31.72, 0.3848; (f) 50, 0.4925, 37.58, 0.9600, 0.9912, 37.58, 0.6839. The four images in each sub figure represents $f$, $Y$, $\hat{f}_{c}$ and $\hat{f}$ respectively. It is clear that FR network performed quite better at high bit-rate \& CR network performed better at low bit-rate. Visually, at low rates, the proposed algorithm is also be seen to reduce blocking artifacts.
The numbers for each sub figure specify the QF, Rate (bpp), PSNR (dB), SSIM, MS-SSIM, PSNRB (dB) and mIoU.}
\label{fig:pic1+pic2}
\vspace{-6mm}
\end{figure*}

\begin{figure*}
\centering


\includegraphics[width=\linewidth,height=6cm]{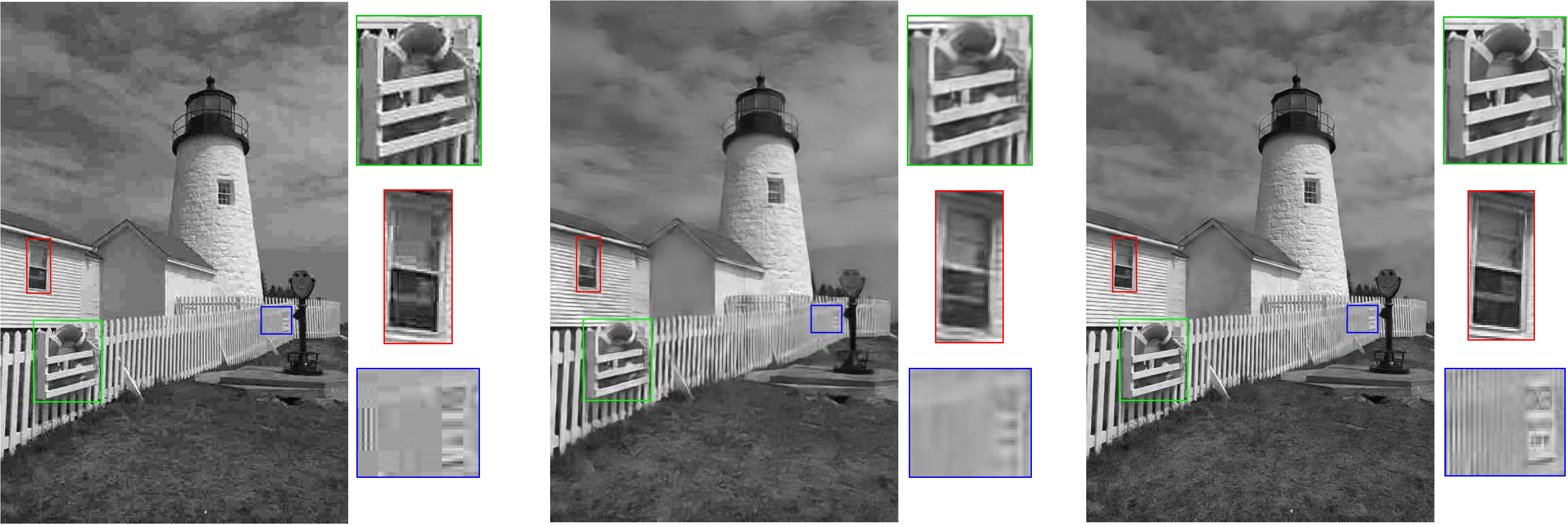}
\caption{ \textit{Left:} At QF:10, JPEG decoded image with bpp=0.1026, PSNR=27.78, SSIM=0.7623, MS-SSIM=0.9182, \textit{Middle:} Post-processed image using proposed CR image representation based algorithm with bpp=0.06, PSNR=25.17, SSIM=0.7546, MS-SSIM=0.8896, \textit{Right:} Post-processed image using proposed FR image representation based algorithm with bpp=0.2074, PSNR=29.97, SSIM=0.8892, MS-SSIM=0.9485. 
Both the CR \& FR image representation network produce images with a quite reduction in blocking artifacts by preserving edges \& textures as compared to JPEG.} 
\label{fig:compare}
\end{figure*}

\begin{table*}[htbp]
\caption{Performance evaluation (PSNR (dB)/SSIM/MS-SSIM) showing the efficacy of HED detector over Canny Edge detector, when tested on Kodak dataset.}
\begin{center}
\begin{tabular}{c|c  |c |c  |c  }
\hline
\hline
\diagbox[width=8em]{Algorithms}{QF}
& \multicolumn{1}{c|}{10} 
& \multicolumn{1}{c|}{20} 
& \multicolumn{1}{c|}{10} 
& \multicolumn{1}{c}{20} 

  \\
\hline
\hline
& \multicolumn{2}{c|}{Canny Edge Detector (CED)} & \multicolumn{2}{c}{HED} \\
\hline

$P_{E} (CR)$ & 24.57/0.7523/0.8946  & 24.80/0.7642/0.9055  &  26.13/0.7679/0.9131  & 27.63/0.8067/0.9483 \\
$P_{E} (FR)$ &  28.99/0.8764/0.9328 & 30.86/0.9077/0.9633  &   30.07/0.8983/0.9574 & 32.24/0.9261/0.9764  \\

\hline
\hline
\end{tabular}
\label{hed}
\end{center}
\vspace{-6mm}
\end{table*}

\begin{table*}[htbp]
\caption{Performance evaluation (PSNR (dB)/SSIM/MS-SSIM) showing the efficacy of EDSR over VDSR post-processing module, when tested on Kodak dataset.}
\begin{center}
\begin{tabular}{c|c  |c |c  |c  }
\hline
\hline
\diagbox[width=8em]{Networks}{QF}
& \multicolumn{1}{c|}{10} 
& \multicolumn{1}{c|}{20} 
& \multicolumn{1}{c|}{30} 
& \multicolumn{1}{c}{40} \\
\hline
\hline
$P_{E_{VDSR}} (CR)$ & 24.61/0.6262/0.8559  & 25.59/0.6800/0.9098 & 26.04/0.7106/0.9304 &  26.35/0.7296/0.9418    \\
$P_{E_{VDSR}} (FR)$ &  27.76/0.7700/0.9290 & 29.79/0.8472/0.9679  & 31.54/0.8795/0.9791 &   32.31/0.8983/0.9844   \\
\hline
$P_{E_{EDSR}} (CR)$ &   26.13/0.7679/0.9131  & 27.63/0.8067/0.9483  & 28.25/0.8304/0.9612  &  28.52/0.8365/0.9658     \\
$P_{E_{EDSR}} (FR)$ &  30.07/0.8983/0.9574 & 32.24/0.9261/0.9764  & 33.55/0.9451/0.9862  &   34.76/0.9507/0.9898   \\

\hline
\hline

\end{tabular}
\label{vdsr}
\end{center}
\vspace{-6mm}
\end{table*}

\begin{table*}[htbp]
\caption{Performance evaluation (PSNR (dB)/SSIM/MS-SSIM) to check the effect of Pre ($P_{r}N$) \& Post Processing Networks ($P_{o}N$), when tested on Kodak dataset.}
\begin{center}
\begin{tabular}{c|c  |c |c  |c  }
\hline
\hline
\diagbox[width=12em]{Networks}{QF}
& \multicolumn{1}{c|}{10} 
& \multicolumn{1}{c|}{20} 
& \multicolumn{1}{c|}{30} 
& \multicolumn{1}{c}{40} \\
\hline
\hline
$P_{r}N + Codec~(CR)$ & 26.88/0.7419/0.9258  & 27.68/0.7996/0.9644  & 27.83/0.8198/0.9758 &  27.90/0.8288/0.9808    \\

$P_{r}N+Codec+P_{o}N~(CR)$ & 26.13/0.7679/0.9131  & 27.63/0.8067/0.9483  & 28.25/0.8304/0.9612  &  28.52/0.8365/0.9658    \\
\hline
$P_{r}N + Codec~(FR)$ & 28.01/0.7702/0.9291  & 30.18/0.8466/0.9678  & 31.60/0.8801/0.9795 &  32.53/0.8990/0.9849    \\
$Codec +P_{o}N~(CR/FR)$ & 28.96/0.7952/0.9413 &  31.22/0.8630/0.9717 & 32.30/0.8900/0.9810  &  33.17/0.9061/0.9856  \\
\hline
$P_{r}N+Codec +P_{o}N~(FR)$ & 30.07/0.8983/0.9574 & 32.24/0.9261/0.9764  & 33.55/0.9451/0.9862  &   34.76/0.9507/0.9898     \\
\hline
\hline
\end{tabular}
\label{prn}
\end{center}
\vspace{-6mm}
\end{table*} 

\begin{table*}[htbp]
\caption{Mean intersection over union showing effectiveness of edge aware loss function over MSE loss function at QF:10 \& 50. $P_{M}$: Proposed algorithm with training of both modules on MSE loss, $P_{E}$: Proposed algorithm with training of $P_{r}N$ \& $P_{o}N$ modules on Edge and MSE loss respectively.}
\begin{center}
\begin{tabular}{c| c c c c c c c c c}
\hline
\hline
\diagbox[width=8em]{Algorithm}{Dataset}  & Set7  & Set14 & Live 1   &   Set5 & Kodak & CLIC 2019 & General 100 & Classic 5 & Average\\
\hline

QF &  \multicolumn{8}{c|}{10} \\
 \hline

\textit{$P_{M}$ (CR)} &  0.3571   &   0.2847 &  0.2503   & 0.3063 & 0.2343  & 0.2496 & 0.2952 &  0.2734 & 0.2814 \\
\textit{$P_{M}$ (FR)} & 0.5775    &  0.5163  & 0.5063     & 0.5105  &  0.4970 & 0.4816  & 0.5373 &  0.5066 & 0.5225\\
\hline
\textit{$P_{E}$ (CR)}& 0.3589  &   0.2815 & 0.2539 & 0.3000  & 0.2375   &  0.2531 & 0.2993  & 0.2739 & 0.2823 \\
\textit{$P_{E}$ (FR)} & 0.6068    & 0.5279   & 0.5322    & 0.5347  & 0.5252   & 0.5020 & 0.5575  &  0.5264 & 0.5391 \\
\hline

QF &  \multicolumn{8}{c}{50} \\
\hline

\textit{$P_{M}$ (CR)}  &  0.5332   &   0.4418 &  0.4006  & 0.4880  & 0.3868  & 0.4185 & 0.4853  & 0.4246 & 0.4474\\
\textit{$P_{M}$ (FR)}  & 0.6413    &  0.5776  & 0.5622    & 0.5393  & 0.5478  &  0.5362 & 0.6093  & 0.5754  & 0.5736  \\
\hline
\textit{$P_{E}$ (CR)}  & 0.5154  &   0.4257 & 0.3900 & 0.4613  & 0.3771  & 0.4041  & 0.4612  & 0.4282 & 0.4329\\
\textit{$P_{E}$ (FR)}  & 0.7676    & 0.7066   & 0.7137    & 0.7134  & 0.7031  & 0.6984 &  0.7393 & 0.7257 & 0.7209 \\
\hline
\hline

\end{tabular}
\label{miou}
\end{center}
\vspace{-6mm}
\end{table*}

\begin{table*}[htbp]
\caption{Performance evaluation (PSNR (dB)/SSIM/MS-SSIM) showing the effectiveness of Edge aware loss function over MSE loss function, when tested on Kodak dataset.}
\begin{center}
\begin{tabular}{c|c  |c |c  |c }
\hline
\hline
\diagbox[width=8em]{Algorithms}{QF}
& \multicolumn{1}{c|}{10} 
& \multicolumn{1}{c|}{20} 
& \multicolumn{1}{c|}{30} 
& \multicolumn{1}{c}{40}\\
\hline
\hline
$P_{M} (CR)$ & 24.30/0.7596/0.8959  &  26.55/0.8048/0.9463 & 27.04/0.8131/0.9542 &  26.65/0.8181/0.9644    \\
$P_{E} (CR)$ & 26.13/0.7679/0.9131  & 27.63/0.8067/0.9483  & 28.25/0.8304/0.9612  &  28.52/0.8365/0.9658     \\
\hline
\hline
$P_{M} (FR)$ & 26.09/0.8887/0.9464  &  31.77/0.8343/0.9585 & 31.86/0.9230/0.9660 & 33.67/0.9415/0.9709     \\
$P_{E} (FR)$ &  30.07/0.8983/0.9574 & 32.24/0.9261/0.9764  & 33.55/0.9451/0.9862  &   34.76/0.9507/0.9898      \\

\hline
\hline
\end{tabular}
\label{mse}
\end{center}
\vspace{-5mm}
\end{table*}

\begin{table}[htbp]
\caption{Performance comparison (PSNR(dB)/SSIM/MS-SSIM) showing effect of CNN layer with stride 2 in $P_{r}N$, when tested on Kodak dataset.}
\begin{center}
\begin{tabular}{c|c  |c   }
\hline
\hline
\diagbox[width=8em]{Stride 2}{QF} & 10 & 20 \\

\hline
At first layer & 26.33/0.7372/0.9095  & 27.55/0.7876/0.9436   \\  
At last layer  & 26.13/0.7679/0.9131  & 27.63/0.8067/0.9483 \\
\hline
\hline
\end{tabular}
\label{stride2}
\end{center}
\vspace{-6mm}
\end{table} 

\begin{figure*}
\centering
\includegraphics[width=\linewidth,height=3.8cm]{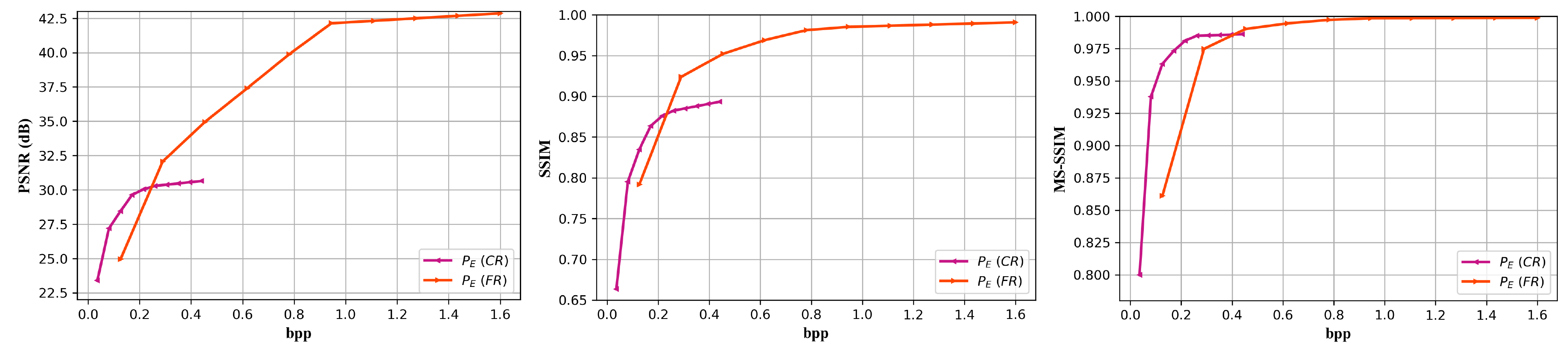}
\caption{Performance comparison showing the applicability of proposed CR \& FR networks based algorithm.}
\label{fig:CR_FR}
\vspace{-4mm}
\end{figure*}

\begin{table*}[htbp]
\caption{Compression performance comparison (PSNR(dB)/SSIM) of proposed algorithm with JPEG \& Jiang\textquotesingle s~\cite{framework}.}
\begin{center}
\begin{tabular}{c| c  c  c | c   c c}
\hline
\hline
QF & \multicolumn{3}{c|}{5} & \multicolumn{3}{c}{10} \\
\hline
\diagbox[width=8em]{Dataset}{Algorithm}
& \multicolumn{1}{c|}{JPEG}
& \multicolumn{1}{c|}{Jiang\textquotesingle s~\cite{framework}} 
& \multicolumn{1}{c|}{$P_{E}$ (FR)}
& \multicolumn{1}{c|}{JPEG}
& \multicolumn{1}{c|}{Jiang\textquotesingle s~\cite{framework}} 
& \multicolumn{1}{c}{$P_{E}$ (FR)}\\
\hline
\hline
Set5 & 26.13/0.7206   & 29.20/0.8387  &   29.98/0.8561  & 28.99/0.8109  & 31.40/0.8854   & 31.67/0.9032\\

Set14 & 24.90/0.6686  & 26.89/0.7914  &  27.00/0.8423 & 27.49/0.7762 & 28.91/0.8336 &  30.21/0.8998 \\

Live 1 & 24.60/0.6666   & 26.78/0.7934  & 26.93/0.8401  & 27.02/0.7720 & 28.84/0.8411  &  29.27/0.9008\\

General 100 & 25.93/0.7228  & 27.29/0.8447  & 28.10/0.8822 &  28.92/0.8199    &  30.16/0.8767 &    31.79/0.9324 \\

\hline
Average & 25.39/0.6947    & 27.54/0.8171     &  28.00/0.8552   &  28.11/0.7948 & 29.83/0.8592   &  30.74/0.9091  \\
\hline
\hline
\end{tabular}
\label{jiang}
\end{center}
\vspace{-8mm}
\end{table*}

\begin{table*}[htbp]
\caption{Comparison (PSNR(dB)/SSIM of the proposed algorithm with other deblocking algorithms at QF:10 as reported in~\cite{framework} on Set 7 dataset (results of other methods have been taken from Jiang\textquotesingle s paper~\cite{framework}).}
\begin{center}
\begin{tabular}{ c|| c c c c c c c||c }
\hline
\hline
Test Images 
& \multicolumn{1}{c}{Lena}
& \multicolumn{1}{c}{Butterfly}
& \multicolumn{1}{c}{Cameraman}
& \multicolumn{1}{c}{House}
& \multicolumn{1}{c}{Peppers}
& \multicolumn{1}{c}{Parrots}
& \multicolumn{1}{c||}{Leaves}
& \multicolumn{1}{c}{Average} \\
\hline\hline
\hline
JPEG~\cite{standard:jpeg94} & 25.24/0.8325 &  26.47/0.7965 &  30.56/0.8183 &  30.41/0.8183 &  30.14/0.7839 &  25.40/0.8609 &  28.96/0.8336 &  28.17/0.8206  \\
Sun\textquotesingle s~\cite{sun} &  26.52/0.8871 &  27.26/0.8358 &  32.00/0.8504 &  31.72/0.8590 &  31.62/0.8322 &  26.60/0.9138 &  30.04/0.8783 &  29.39/0.8652  \\
Foi\textquotesingle s~\cite{foi} & 31.84/0.8586 & 27.25/0.9014  & 27.480.8333  &  32.09/0.8491 &  31.69/0.8318 &  30.15/0.8778 & 27.30/0.9282 &  29.69/0.8686    \\
BM3D~\cite{bm3d} & 26.64/0.8896 &  27.25/0.8240 &  32.07/0.8492 &  31.77/0.8549 &  31.42/0.8250 &  26.98/0.9207 &  30.05/0.8749 &  29.45/0.8626    \\
Zhang\textquotesingle s~\cite{overlap} &  26.83/0.8923 &  27.45/0.8329 &  32.11/0.8513 &  31.92/0.8597 &  31.68/0.8317 &  27.26/0.9212 &  30.50/0.8804 &  29.68/0.8671\\
Ren\textquotesingle s~\cite{ren} & 27.17/0.9010 &  27.43/0.8259 &  32.41/0.8526 &  31.92/0.8571 &  31.63/0.8300 &  27.59/0.9309 &  30.34/0.8775 &  29.78/0.8679    \\
DicTV~\cite{dictv} & 26.09/0.8699 &  26.92/0.8046 &  31.77/0.8484 &  31.55/0.8559 &  31.29/0.8244 &  26.33/0.9032&  29.82/0.8741 &  29.11/0.8544     \\
WNNM~\cite{wnnm}&  27.22/0.9019 &  27.40/0.8248 &  32.42/0.8531 &  31.93/0.8571 &  31.64/0.8303 &  27.66/0.9325 &  30.33/0.8755 &  29.80/0.8681   \\
CONCOLOR~\cite{concolor} &  27.09/0.9142 &  27.72/0.8401 &  33.04/0.8609 &  32.19/0.8661 &  31.94/0.8358 &  28.20/0.9406 &  30.66/0.8842 &  30.12/0.8774   \\
ARCNN~\cite{arcnn} &  28.54/0.9237 &  27.62/0.8389 &  32.53/0.8591 &  32.0/0.8711 &  31.50/0.8434 &  28.31/0.9495 &  30.62/0.8942 &  30.16/0.8828   \\
ComCNN~\cite{framework} &  26.32/0.8796 &  27.05/0.8154 &  31.82/0.8497 &  31.63/0.8573 &  31.04/0.8296 &  26.51/0.9095 &  29.97/0.8766 &  29.12/0.8597  \\
RecCNN~\cite{framework} &  28.04/0.9192 &  27.33/0.8429 &  32.76/0.8584 &  32.35/0.8679 &  31.34/0.8396 &  28.53/0.9443 &  30.85/0.8884 &  30.17/0.8801  \\
Jiang\textquotesingle s~\cite{framework} &   28.60/0.9245 &  27.44/0.8448 & 33.25/0.8678 &  33.11/0.8838  &  31.83/0.8532 &  28.77/0.9475 &  31.11/0.9047 &  30.59/0.8895 \\
\hline
& \multicolumn{7}{c||}{FR} \\
\hline
$P_{r}N$ & 29.99/0.7938 & 25.37/0.8600 & 26.46/0.7956 & 30.50/0.8187 & 30.10/0.7834 &  31.83/0.8527 & 25.23/0.8245  & 28.50/0.8184  \\
\hline
$P_{o}N$ & 31.19/0.8264 &  27.30/0.9190 &  27.65/0.8281 &  31.91/0.8465 & 31.47/0.8259  & 33.25/0.8912  &  27.53/0.8920 & 30.04/0.8613 \\
\textit{$P_{E}$ (FR)} &  32.28/0.8862  &  28.52/0.9598 & 29.13/0.9080   & 31.95/0.9078  & 31.97/0.8698  &  33.78/0.9361 &   29.06/0.9399 &  30.96/0.9154   \\
\hline
\hline

\end{tabular}
\label{set7}
\end{center}
\vspace{-4mm}
\end{table*}

\begin{table}[htbp]
\caption{PSNR(dB)/SSIM comparison of various deblocking \& post-processing methods with the proposed algorithm on Classic 5 dataset.}
\begin{center}
\begin{tabular}{c|c  | c| c | c}
\hline
\hline
\diagbox[width=8em]{Algorithm}{QF}
& \multicolumn{1}{c|}{10} 
& \multicolumn{1}{c|}{20} 
& \multicolumn{1}{c|}{30} 
& \multicolumn{1}{c}{40}  \\
\hline
\hline
DnCNN-3~\cite{dcnn} &  29.40/0.8026 &  31.63/0.8610  &   32.90/0.8860 &   33.77/0.9003   \\
ARCNN~\cite{arcnn} &  29.05/0.7929 &  31.16/0.8517 & 32.52/0.8806 &   33.33/0.8953     \\
DPW-SDNet~\cite{dpw} &  29.74/0.8124 &  31.95/0.8663  & 33.22/0.8903  &  34.07/0.9039   \\
\hline
\textit{$P_{E}$ (FR)} &   30.18/0.8817 &  32.26/0.9148  &   33.90/0.9357  &   34.59/0.9405     \\
\hline
\hline
\end{tabular}
\label{classic}
\end{center}
\vspace{-8mm}
\end{table}

\begin{table}[htbp]
\caption{PSNR(dB)/SSIM comparison of various deblocking \& post-processing methods with the proposed algorithm on Live 1 dataset.}
\begin{center}
\begin{tabular}{c|c  |c |c  |c }
\hline
\hline
\diagbox[width=8em]{Algorithm}{QF}
& \multicolumn{1}{c|}{10} 
& \multicolumn{1}{c|}{20} 
& \multicolumn{1}{c|}{30} 
& \multicolumn{1}{c}{40}  \\
\hline
\hline
DnCNN-3~\cite{dcnn} &  29.19/0.8123 &  31.59/0.8802  &   32.98/0.9090 &   33.96/0.9247    \\
ARCNN~\cite{arcnn} &  28.98/0.8217 &  31.29/0.8871 & 32.69/0.9166 & 33.63/0.9306     \\
DPW-SDNet~\cite{dpw} &  29.53/0.8210 & 31.90/0.8854  & 33.31/0.9130  &  34.30/0.9282   \\
CASCNN~\cite{cascnn}\tablefootnote{The results of Cavigelli et al.~\cite{cascnn} are not given at QF=30.} & 29.44/0.8330 &  31.70/0.8950 & - & 34.10/0.9370 \\
Galteri\textquotesingle s~\cite{galteri} & 29.45/0.8340 & 31.77/0.8960 & 33.15/0.9220 & 34.09/0.9350 \\

\hline
\textit{$P_{E}$ (FR)} & 29.27/0.9008   & 30.90/0.9288  &  32.74/0.9475  &   33.93/0.9535    \\
\hline
\hline
\end{tabular}
\label{live}
\end{center}
\vspace{-8mm}
\end{table}

\subsection{Evaluation Parameters}
The original image $\mathbf{f}(x,y)$ has been compared with $\hat{\mathbf{f}}(x,y)$ for the assessment of the proposed compression-decompression algorithm.
We have used both objective and subjective performance metrics i.e., PSNR and SSIM for a comprehensive assessment. Along with the consideration of structural information, a metric should be chosen which incorporates image details at various resolutions. Therefore, MS-SSIM, a measure based on structural information at different scales is chosen for assessment.
In the same direction, we have also calculated PSNRB~\cite{psnrb} which incorporates the effect of blocking artifacts (by incorporating the blocking effect factor) while calculating the PSNR. To check for the edge-preserving property of edge-aware loss function we have calculated intersection over union (IoU) to check the overlap regions between true edges and retained edges. 
Bjontegaard metric calculation~\cite{bd} has also been conducted to compare the coding gain achieved with the proposed algorithm with the other methods. The BD metric facilitate the comparison of area under rate-distortion curves in terms of the average PSNR improvement or the average percent bit-rate saving. A negative (positive) value of BD-Rate indicates a decrease (increase) of bit-rate for the same PSNR. A positive (negative) value of BD-PSNR indicates an increase (decrease) of PSNR for the same bit-rate.

\begin{table}[htbp]
\caption{PSNRB(dB) comparision of various deblocking \& post-processing methods with the proposed algorithm on Live 1 dataset.}
\begin{center}
\begin{tabular}{c|c  |c |c  |c }
\hline
\hline
\multicolumn{1}{c|}{QF}
& \multicolumn{1}{c|}{10} 
& \multicolumn{1}{c|}{20} 
& \multicolumn{1}{c|}{30} 
& \multicolumn{1}{c}{40}  \\

\hline
\hline
ARCNN~\cite{arcnn} &  28.77 & 30.79 &  32.22 & 33.14   \\
CASCNN~\cite{cascnn} & 29.19 & 30.88 & - & 33.68 \\
DnCNN-3~\cite{dcnn} &  28.91 & 31.08 & 32.35 & 33.29    \\
Galteri\textquotesingle s~\cite{galteri} & 29.10 & 31.26 & 32.51 & 33.40 \\
DPW-SDNet~\cite{dpw} &  29.13 & 31.27 & 32.52 & 33.44  \\

\hline
\textit{$P_{E}$ (FR)} & 29.27   & 30.89  & 32.74  &  33.92    \\
\hline
\hline
\end{tabular}
\label{psnrblive1}
\end{center}
\vspace{-4mm}
\end{table}

\begin{table}[htbp]
\caption{PSNRB(dB) comparision of various deblocking \& post-processing methods with the proposed algorithm on Classic 5 dataset.}
\begin{center}
\begin{tabular}{c|c  |c |c  |c }
\hline
\hline
\multicolumn{1}{c|}{QF}
& \multicolumn{1}{c|}{10} 
& \multicolumn{1}{c|}{20} 
& \multicolumn{1}{c|}{30} 
& \multicolumn{1}{c}{40}  \\
\hline
\hline


ARCNN~\cite{arcnn} & 28.78 & 30.60 & 32 & 32.81    \\
DnCNN-3~\cite{dcnn} &  29.13 & 31.19 & 32.26 & 33.20    \\
DPW-SDNet~\cite{dpw} &   29.37 & 31.42 & 32.51 & 33.24 \\
\hline
\textit{$P_{E}$ (FR)} &  30.14  &  32.21 & 33.83   & 34.51     \\
\hline
\hline

\end{tabular}
\label{psnrbclassic5}
\end{center}
\vspace{-6mm}
\end{table}


\begin{table}[htbp]
\caption{BD-Rate (in \%)/BD-PSNR (dB) for coding gain obtained comparing proposed algorithm with-recent state-of-the art techniques.}
\begin{center}
\begin{tabular}{c|c  |c  }
\hline
\hline
\diagbox[width=12em]{Other Algo}{Proposed}& $P_{E}~(CR)$ & $P_{E}~(FR)$ \\ 
\hline
\multicolumn{3}{c}{Set 5}  \\
\hline
JPEG  &  -68.04/5.71 &  -56.67/5.96   \\
JPEG2000    & -44.99/2.67 & -33.54/2.71 \\
Zhao\textquotesingle s~\cite{virtual}   & -44.76/2.97 &  -33.72/2.85  \\
\hline
\multicolumn{3}{c}{Set 7}  \\
\hline
JPEG  & -69.81/5.43 & -58.98/6.01     \\
JPEG2000    & -42.96/2.30 &  -30.76/2.39 \\
Zhao\textquotesingle s~\cite{virtual}  & -48.12/2.94 &  -33.94/2.73   \\
\hline
\multicolumn{3}{c}{Set 14}  \\
\hline
JPEG  & -66.59/4.45 &  -57.15/5.70  \\
JPEG2000    & -33.72/1.41 &  -29.00/2.35\\
Zhao\textquotesingle s~\cite{virtual}  & -35.25/1.51 &  -29.50/2.39    \\
\hline
\multicolumn{3}{c}{Live 1}  \\
\hline
JPEG  & -61.63/3.22 &  -49.80/4.75   \\
JPEG2000    & -20.70/0.4997 &  -20.16/1.61 \\
Zhao\textquotesingle s~\cite{virtual}  & -43.84/1.96 &  -36.47/3.20    \\
\hline
\multicolumn{3}{c}{Kodak}  \\
\hline
JPEG  & -69.44/3.07 & -64.18/6.97    \\
JPEG2000    &  -29.87/0.93 & -42.00/3.00 \\
BPG & -3.98/-0.12 &  -14.85/1.65   \\
Balle\textquotesingle s~\cite{optimized}              &  -30.69/0.95 & -13.33/1.88      \\
Theis\textquotesingle s~\cite{theis}                  &    -57.74/2.31 &  -43.94/4.16 \\
Balle\textquotesingle s~(MSE opt.)~\cite{balle2}      &   -25.83/0.49 & -26.05/1.97     \\
Balle\textquotesingle s~(MS-SSIM opt.)~\cite{balle2}  &   -62.19/2.77 & -57.87/5.43     \\
Lee\textquotesingle s~(MSE opt.)~\cite{lee}           &  2.52/-0.43  &    -19.85/1.36   \\
Lee\textquotesingle s~(MS-SSIM opt.)~\cite{lee}       &    -63.97/2.55 &  -61.10/5.88    \\
Choi\textquotesingle s~\cite{choi}                    &  3.44/-0.42 & -16.89/1.28        \\
\hline
\multicolumn{3}{c}{CLIC 2019}  \\
\hline
JPEG2000  & -71.26/4.51 &  -54.22/5.07   \\
BPG   & -59.19/3.14 & -44.14/3.55  \\
Balle\textquotesingle s~\cite{optimized}              & -68.87/4.23 &   -44.36/4.24   \\
Balle\textquotesingle s~\cite{balle2}                 & -64.51/3.70 &  -49.74/4.33   \\
Yang\textquotesingle s~\cite{yang}                    &  -39.78/1.72 & -22.09/1.81   \\
\hline
\hline
\end{tabular}
\label{bd}
\end{center}
\vspace{-6mm}
\end{table}

\section{Ablation Studies} \label{sec:ablation}
The proposed network has been trained separately for CR \& FR representation for a wide range of QFs, i.e., 2, 5, 6, 10, 20, 30, 40, 50, 60, 80, 90 \& 100. In the sections below, we justify the formulation of the proposed algorithm based on various aspects/choices available.

\subsection{Comparison of training the pre-processing network separately with Canny edge detector (CED) \& learning-based detector (HED)}
Many edge detection algorithms are available in the literature. CED is one of the good classical edge detection algorithms which uses the idea of change in intensity on each pixel in an image, i.e., the intensity is very high on the edges. CED identifies edges in 4 steps: noise removal, gradient calculation, non-maximal suppression, and hysteresis thresholding.
Since the CED only focuses on local changes and it has no semantic understanding (understanding the content of the image), it has limited accuracy. Semantic understanding is crucial for edge detection that is why learning-based detectors generate better results than CED. Hence, we have separately tested the classical Canny Edge Detection~\cite{canny} and learning-based edge detector. For learning-based edge detection, we have chosen the Holistically-Nested Edge Detection technique (HED), which has been set as a benchmark technique for edge detection after the development in the CNN architectures. 
HED is a learning-based end-to-end edge detection system that uses a trimmed VGG-like CNN for an image to image prediction task.
It is based on multi-scale and multi-level feature learning and claims to learn rich hierarchical features which are crucial. However, Canny edges are not directly connected and produce some spatial shifts and inconsistencies. 
It makes use of the side outputs of intermediate layers to learn edges synchronizing with the contextual information leaving undesired and false edges. 
Since the feature maps generated at each layer are of different sizes, it effectively looks at images at different scales. Generally, edge detectors are supposed to assume 90\% of the ground truth as non-edge, but HED uses a class balancing weight to make a balance between edge \& non-edge region, to find the true \& strong edges much effectively as shown in Fig.~\ref{fig:hed}. To implement it, we have used the code provided at \footnote{\url{https://github.com/senliuy/Keras_HED_with_model}} for exploiting edge map predictions obtained through training the modified VGGNet with BSDS500~\cite{bsds500}. HED seems to produce better compression performance than CED as shown in Table~\ref{hed}.

\subsection{Comparison for learning the proposed network separately with VDSR \& EDSR post-processing module}
In~\cite{framework}, the VDSR module has been used as the post-processing module. However, we have used the EDSR module to remove the blocking effects generated after JPEG decoding. To show the efficacy of the EDSR module, we have implemented the proposed algorithm considering the VDSR post-processing module. The compression with the EDSR module achieves significant improvement over compression with the VDSR module as shown in Table~\ref{vdsr}.

\begin{figure}
\centering



\includegraphics[width=0.5\textwidth,height=17cm]{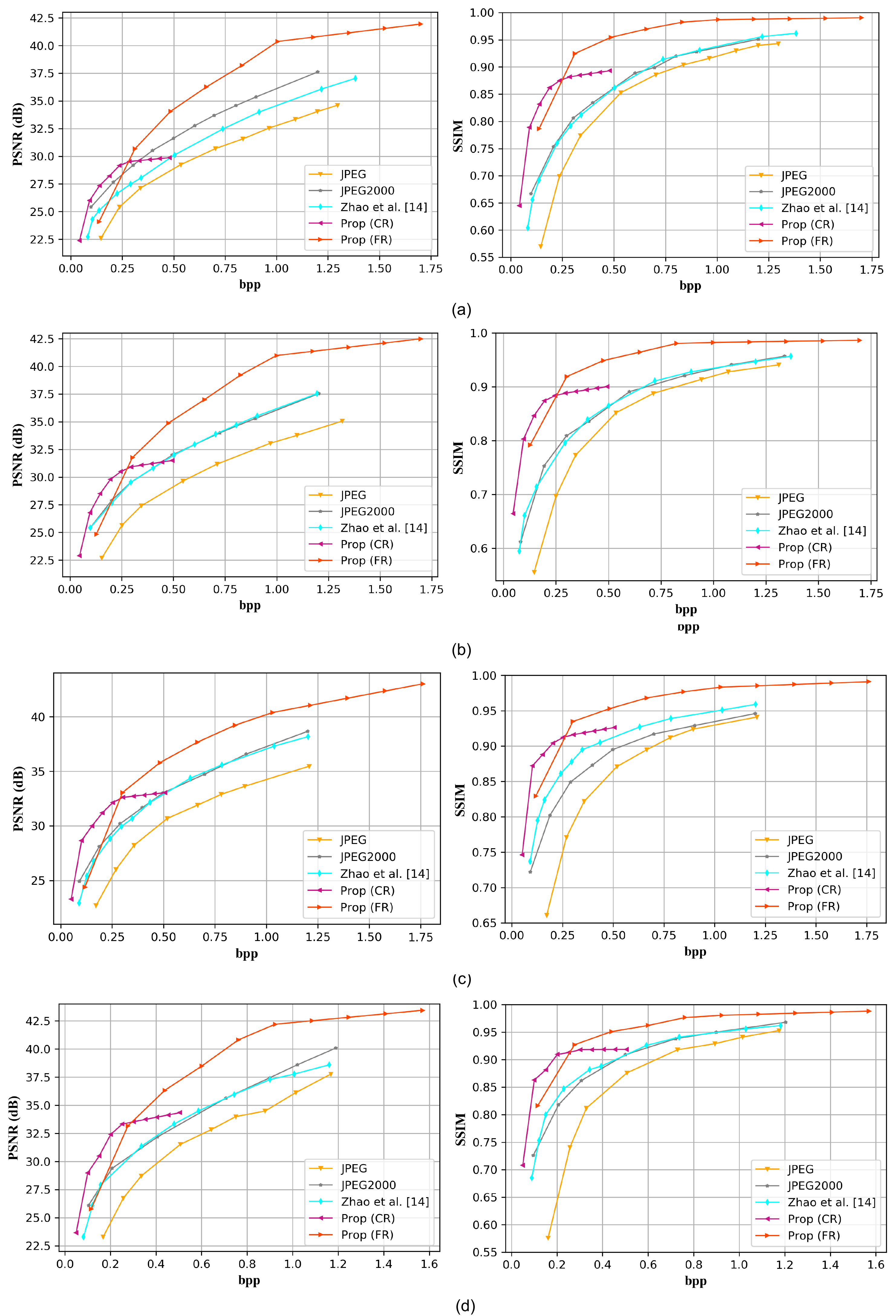}
\caption{Objective (PSNR) \& Subjective parameter (SSIM) comparison of proposed algorithm with standard \& downsampling method on (a) Live 1 (b) Set 14 (c) Set 7 (d) Set 5 datasets.} 
\label{fig:rd_measure}
\vspace{-4mm}
\end{figure}

\begin{figure*}
\centering
\includegraphics[width=\linewidth,height=6cm]{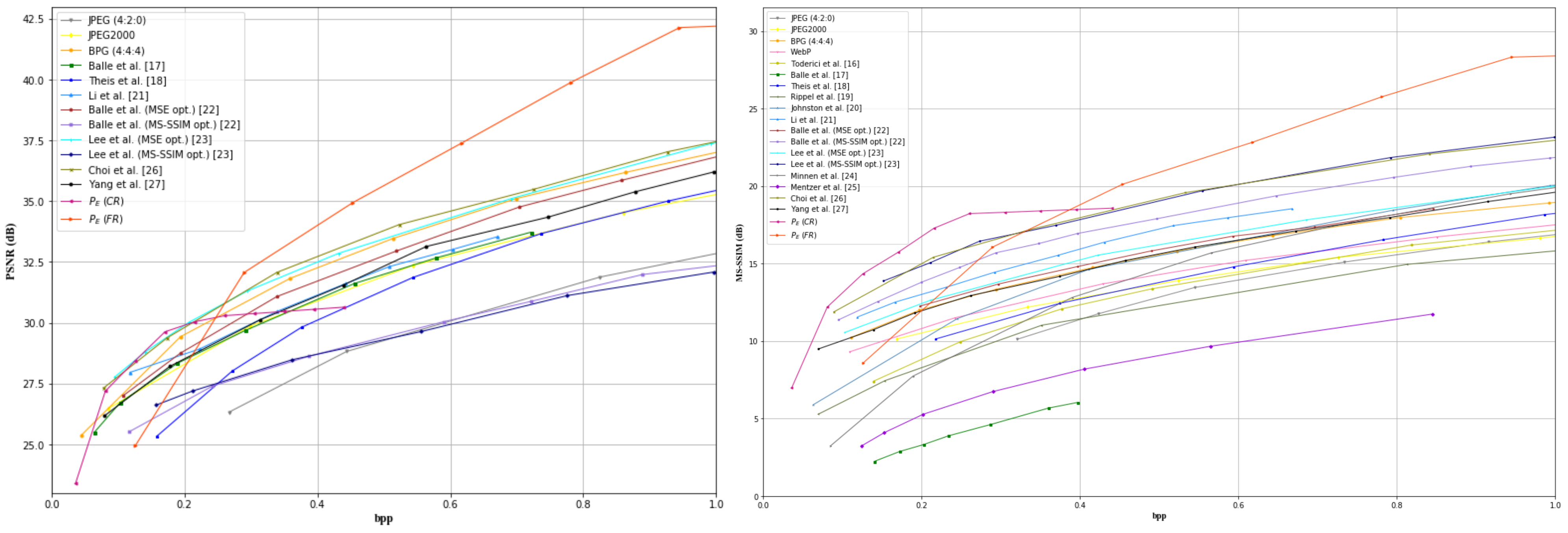}
\caption{Objective (PSNR) \& Subjective parameter (MS-SSIM) comparison of proposed algorithm with standard \& recent state-of-the-art approaches on Kodak dataset. For showing comparison, $-10log_{10}(MS-SSIM)$ is plotted in dB to synchronize with the process adopted in recent state-of-the-art methods.}
\label{fig:kodak_rd_measure}
\end{figure*}

\begin{figure*}
\centering
\includegraphics[width=\linewidth,height=6cm]{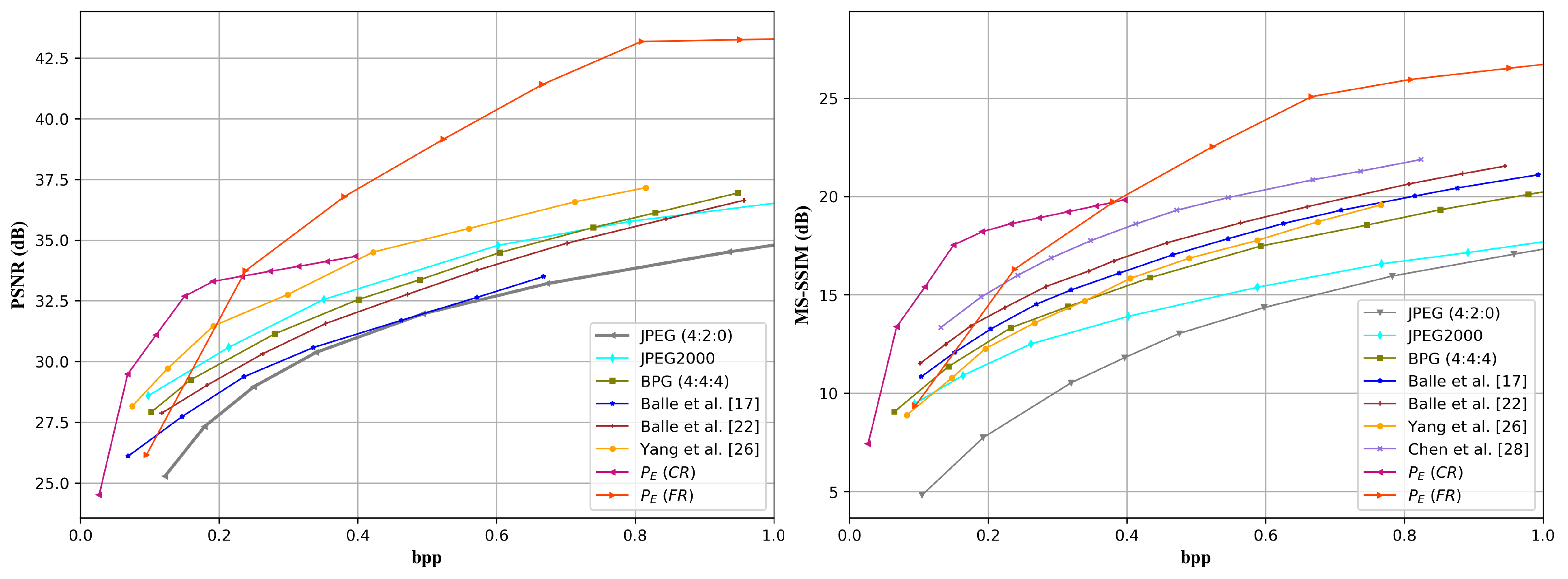}
\caption{Objective (PSNR) \& Subjective parameter (MS-SSIM) comparison of proposed algorithm with standard \& recent state-of-the-art approaches on CLIC 2019 dataset.}
\label{fig:clic_rd_measure}
\end{figure*}

\subsection{Comparison for impact of Pre \& Post-processing module separately on codec compression with full framework}
To check the individual contribution of each sub-module i.e., pre \& post-processing module, we have trained the network considering $P_{r}N$ \& $P_{o}N$ individually. In fact, these individual implementations are the special cases of proposed $P_{E}~(FR)$. From Table~\ref{prn}, it is inferred that traditional codec compression followed by the post-processing network $P_{o}N$ produces better results than pre-processing module $P_{r}N$. Moreover, adding both these modules with the codec compression added further gain in compression performance.

\subsection{Comparison for learning the network on Edge aware loss \& comparison with the MSE loss function}
The training of $P_{r}N$ on edge-aware loss forced the network to retain structural and sharp details while compressing the images. Also, the tuning parameter $\alpha=0.75$ worked best for training the network to preserve edges in the reconstructed images. The first term in Eq.~\eqref{eq:lr} ensures high PSNR by minimizing MSE while preserving edges. Preserving edges via the second term in Eq.~\eqref{eq:lr} ensures the preservation of structural information and hence maximize SSIM \& MS-SSIM.
High mIoU value in Table~\ref{miou} indicate to retain edges with edge loss (in $P_{E}$) as compared to MSE loss function (in $P_{M}$), ultimately ensuring the edge-preserving property. Also, it indicates that an FR network can recover images with quite a higher mIoU than a CR representation network. In addition to this, we have trained both the modules with MSE and the results are given in Table~\ref{mse}. The results show that training of $P_{r}N$ with edge-aware loss function leads to better results than training with MSE. Apart from that, Li et al.~\cite{feng} use a resolution decrement parameter (layer with stride 2) at the first CNN layer. 
But, we believe it should be placed at the last CNN layer (as used in the proposed algorithm) because gradual or delayed downsampling help in learning more features as compared to sudden downsampling just after the first layer. This sudden downsampling may lose learning of some features which are crucially required. To support this statement, a small experiment is conducted first by taking stride 2 in the first layer, then in last and the results are shown in Table~\ref{stride2}.
With multiple sets of experiments, we finalize the training of $P_{r}N$ \& $P_{o}N$ module on edge-aware \& MSE loss functions, respectively. It was found that training with edge-aware loss functions performed well with comparatively more bit saving. 

\subsection{Comparison of CR learning framework with FR learning framework}
The results in Fig.~\ref{fig:pic1+pic2} stressed upon the point of more appropriate learning of features showing the superiority of the FR over CR image representation in terms of compression gain and performance measures. From the highlighted region in the image given in Fig.~\ref{fig:compare}, it is quite evident that at low bit-rates, the proposed $P_{E}~(CR)$ \& $P_{E}~(FR)$ reduces the blocking artifacts more effectively compared to JPEG, due to super-resolution network introduced to post-process the image. 
However, it is noted that better performance at low bit-rates using CR network representation is obtained at the cost of performance saturation at high bit-rates which can be seen from Fig.~\ref{fig:CR_FR}.

\section{Comparison with State-of-the-Art algorithms}
\label{sec:comparison}
 We have compared the proposed algorithm with JPEG and Jiang\textquotesingle s~\cite{framework} for QF of 5 \& 10 on Set 5, Set 14, Live 1 \& General 100 dataset. It is observed that the proposed $P_{E}~(FR)$ has outperformed the mentioned approaches with a significant margin as shown in Table~\ref{jiang}. Specifically, we have compared the proposed algorithm with various deblocking and post-processing based algorithms and again found $P_{E}~(FR)$ to overcome the approaches shown in Table~\ref{set7}. Here also, the contribution of each module $P_{r}N$ \& $P_{o}N$ can be seen as demonstrated separately.

When tested on Classic 5 \& Live 1 dataset, the proposed algorithm clearly outperformed the work reported in~\cite{dcnn,arcnn} as shown in Table~\ref{classic} \& Table~\ref{live} respectively. The proposed algorithm is effectively seen as to remove blocking artifacts which are confirmed when comparing PSNRB with standard artifacts reduction techniques as shown in Table~\ref{psnrblive1} \& Table~\ref{psnrbclassic5}. 
\begin{table*}[htbp]
\caption{Similarity tests (average value) between training set (BSD400) and testing sets used.}
\begin{center}
\begin{tabular}{ c| c | c | c |c |c  |c |c | c}
\hline
\hline
Datasets &  Set 5 & General 100  & Set 7 & CLIC 2019 & Set 14 & Kodak & Live 1 & Classic 5  \\
\hline
Correlation Coefficient & 0.0488 & 0.0985 & 0.109 &  0.1106 & 0.1164 & 0.126  & 0.1521 & 0.2475   \\
\hline
\hline
\end{tabular}
\label{test}
\end{center}
\vspace{-4mm}
\end{table*}

\begin{table}[htbp]
\caption{Run time complexity (in sec) of proposed algorithm with various state-of-the-art algorithms for compression of $256\times256$ size image at QF:10.}
\begin{center}
\begin{tabular}{c c| c c }
\hline
\hline
Algorithm & Time (CPU/GPU) & Algorithm & Time (CPU/GPU) \\
\hline
ARCNN~\cite{arcnn} & -/0.33 & Zhang\textquotesingle s~\cite{concolor} &  442/-  \\
Jiang\textquotesingle s~\cite{framework} & 1.56/0.017 & DnCNN-3~\cite{dcnn} & -/0.0157\\
Sun\textquotesingle s~\cite{sun} & 132/- & Rippel\textquotesingle s~\cite{rippel} & -/0.01 \\
BM3D~\cite{bm3d} & 3.40/-  & Balle\textquotesingle s~\cite{balle2} & 0.7/- \\
Zhang\textquotesingle s~\cite{overlap} & 251/-   & Lee\textquotesingle s~\cite{lee} & -/7.5 \\
Ren\textquotesingle s~\cite{ren} &  36/-  & DPW-SDNet~\cite{dpw} &  -/$1.2\sim2$ \\
DicTV~\cite{dictv} &  53/- &    CASCNN~\cite{cascnn} & -/2.4  \\
WNNM~\cite{wnnm}  &  230/-  & \textit{Proposed CR/FR} &  -/0.081\\
\hline
\hline
\end{tabular}
\label{time}
\end{center}
\vspace{-2mm}
\end{table}

\begin{table}[htbp]
\caption{Training parameters or model size comparison of proposed algorithm with various state-of-the-art methods.}
\begin{center}
\begin{tabular}{ c c | c c }
\hline
\hline
Algorithm &  Parameters & Algorithm &  Parameters \\
\hline

ARCNN~\cite{arcnn} & 106564 (417 KB) &  Li et al.~\cite{learning} &     257557187  \\
Jiang\textquotesingle s~\cite{framework} & 708674  &  Balle\textquotesingle s~\cite{balle2} & 4.76 million     \\
Zhao\textquotesingle s~\cite{virtual} & 19968768  & Mentzer\textquotesingle s~\cite{mentzer} & 9561472    \\
DnCNN-3~\cite{dcnn} & 1112192 & Yang\textquotesingle s~\cite{yang} & 10.27 million   \\
Toderici\textquotesingle s~\cite{fullresolution} &    7457472  & Chen\textquotesingle s~\cite{chen} & 261.803 MB   \\
Balle\textquotesingle s~\cite{optimized} &    16.1 million &   DPW-SDNet~\cite{dpw}  &  1700736    \\
Theis\textquotesingle s~\cite{theis} & 2742112  & CASCNN~\cite{cascnn} & 5144000 \\
Johnston\textquotesingle s~\cite{johnston} & 9917504 &   Galteri\textquotesingle s~\cite{galteri} & 7128160   \\
\hline
\textit{Proposed CR/FR} & 796098/648386  \\


\hline
\hline
\end{tabular}
\label{tabparameter}
\end{center}
\vspace{-4mm}
\end{table}

When tested with Live 1, Set 14, Set 7 \& Set 5 dataset, the proposed algorithm has effectively outperformed JPEG, JPEG2000 \& Zhao\textquotesingle s~\cite{virtual} as shown in Fig.~\ref{fig:rd_measure}). For plotting rate-distortion curves, the bit-rate for a file is obtained by dividing the file size of codec compressed bit-stream with the respective original image dimension.
With the proposed algorithm, the improvement in PSNR is approximately 25\%, 3.77\%, 10\%, and 22.73\%, 12.5\%, 19.12\% at low and high bit-rates respectively, compared to JPEG, JPEG2000, and Jiang\textquotesingle s algorithms on Live 1 dataset. This improvement in SSIM is approximately 45.6\%, 15.28\%, 16.9\%, and 10.11\%, 6.5\%, 6.52\% at low and high bit-rates respectively, compared with the same approaches. 
When tested on the Set 14 dataset, the improvement in PSNR is 26.67\%, 9.61\%, 9.61\%, and 20.31\%, 13.24\%, 11.59\% at low and high bit-rates respectively, compared to JPEG, JPEG2000 and Jiang\textquotesingle s algorithms. Similarly, the increase in SSIM is approximately 41.67\%, 16.44\%, 19.72\%, and 11.90\%, 6.67\%, 5.49\% at low and high bit-rates respectively, compared with the same approaches.
On the Set 7 dataset, the improvement in PSNR is approximately 30.43\%, 11.11\%, 15.38\%, and 16.92\%, 7.04\%, 8.57\% at low and high bit-rates respectively, compared to JPEG, JPEG2000 and Jiang\textquotesingle s algorithms. Similarly, the improvement in SSIM is 31.82\%, 12.99\%, 8.75\%, and 6.59\%, 5.43\%, 3.19\% at low and high bit-rates respectively, compared with the same approaches.
The proposed algorithm outperformed JPEG, JPEG2000, and Jiang\textquotesingle s with PSNR with an approximate gain of 38.30\%, 12.07\%, 16.07\%, and 19.40\%, 12.68\%, 12.68\% at low and high bit-rates respectively, on Set 5 dataset. The improvement in SSIM is approximately 42.59\%, 3.90\%, 7.3\%, and 7.87\%, 3.23\%, 2.13\% at low and high bit-rates respectively, compared with the same approaches.
As shown in Fig.~\ref{fig:kodak_rd_measure}, when tested on Kodak benchmark dataset, the proposed algorithm outperformed JPEG, JPEG2000, BPG, and the work reported in~\cite{optimized,theis,learning,balle2,lee,choi,yang} in terms of PSNR by an approximate improvement of 20.75\%, 8.47\%, 3.22\%, 8.47\%, 14.29\%, 4.17\%, 4.92\%, 3.23\%, 3.23\%, 5.26\% and 24.59\%, 14.46\%, 10.14\%, 13.43\%, 16\%, 13.64\%, 11.76\%, 8.57\%, 8.57\%, 14.94\% at low and high bit-rates respectively. Similarly the improvement in MS-SSIM is approximately 71.43\%, 50\%, 36.36\%, 23.08\%, 56.52\%, 40.63\%, 63.64\%, 50\%, 70\%, 25.93\%, 9.09\%, 6.5\%, 13.04\%, 67\%, 64.70\%, 41.67\% and 64.47\%, 61.29\%, 47.06\%, 51.52\%, 60.26\%, 43.68\%, 78.57\%, 51.52\%, 32.35\%, 27.78\%, 21.95\%, 35.14\%, 42.86\%, 69\%, 16.28\%, 35.29\% at low and high bit-rates respectively, when compared with JPEG, JPEG2000, BPG, WebP algorithms and the work in~\cite{fullresolution,optimized,theis,rippel,johnston,learning,balle2,lee,minnen,mentzer,choi,yang}. 

When tested on CLIC 2019 challenging dataset as shown in Fig.~\ref{fig:clic_rd_measure}, the proposed algorithm outperformed JPEG, JPEG2000, BPG and the work in~\cite{optimized,balle2,yang} by approximately 20.73\%, 8.85\%, 10.53\%, 16.24\%, 11.70\%, 6.78\% and 23.08\%, 14.94\%, 17.39\%, 22.73\%, 19.12\%, 14.08\% at low and high bit-rates respectively. Similarly the improvement in MS-SSIM is approximately 72\%, 45.45\%, 39.13\%, 33.33\%, 23.08\%, 52.38\%, 18.52\% and 71.43\%, 50\%, 41.18\%, 33.33\%, 29.73\%, 37.14\%, 17.07\% at low and high bit-rates respectively, when compared to JPEG, JPEG2000, BPG and the work in~\cite{optimized,balle2,yang,chen}.
More gain in terms of objective (PSNR) \& subjective (MS-SSIM) evaluation has been achieved at lower as well as higher bit-rates due to the use of a super-resolution network that uses residual learning. The edge-aware loss function also helped to reduce the bit-rate while maintaining the same quality of decompressed images. The coding gain of the proposed algorithm (BD metric calculation) concerning various state-of-the-art methods is also found to be quite high as given in Table~\ref{bd}. High coding gain means that a very large amount of bit-saving takes place on the application of proposed compression-decompression algorithm. 
Comprehensively, it can be deduced that the objective (PSNR) and perceptual subjective quality metric (SSIM \& MS-SSIM), shows significant improvement with compression rates.
During the investigation of the results obtained, we have also conducted a correlation analysis to check the similarity/dissimilarity between training and testing sets in Table~\ref{test}.  
For this, we have calculated the correlation coefficient as a metric, which seems to synchronize with the compression performance obtained in rate-distortion curves. From Fig.~\ref{fig:rd_measure}, Fig.~\ref{fig:kodak_rd_measure}, Fig.~\ref{fig:clic_rd_measure} \& Table~\ref{test}, it is clear that if the correlation coefficient between training and all testing sets is more, better compression performance will be obtained on that test set \& vice-versa.
Through the extensive investigation, we report that, $P_{E}~(CR)$ \& $P_{E}~(FR)$ representation network is suitable for low \& high bit-rates respectively. The compression performance is saturated at high bit-rates when using the CR representation network. Since it is a well-known fact that to display images on Ultra high definition (UHD), initial downsampling of images is mandatory. So compact-resolution image compression $P_{E}~(CR)$ can be preferred where transmission bandwidth is limited \& vice-versa.
Furthermore, training with a large size dataset did not help in improving the efficiency of the compression algorithm, which is also supported by~\cite{framework}. Compared with standard approaches, deblocking methods, post-processing methods, recently reported state-of-the-art techniques, the approach yields very good results with only a few training images. The testing time when using the proposed algorithm is also less i.e., 0.081s, making it practically implementable.
Table~\ref{time} and Table~\ref{tabparameter} show the comparison between the run-time complexity and number of training parameters for the proposed algorithm. Both the testing time and training parameters being feasible to make the proposed algorithm real-time applicable.

\section{Conclusion} \label{sec:conclusion}
The end-to-end compression-decompression framework centered on super-resolution preserves the quality of the images both at low \& high bit-rates along with reducing artifacts while reconstructing the images. The edge-aware loss function introduced using HED is used to prevent the blurriness \& artifacts which are generally caused by training the network with MSE. We reported on extensive experiments conducted to evaluate the proposed approach on several compression datasets. The results demonstrate that compared to the various state-of-the-art method, both the super-resolution post-processing network \& the edge-aware loss function provide significant gain concerning deblocking, rate-distortion performance, compression performance (PSNR, PSNRB, SSIM, MS-SSIM), and coding gain in terms of BD-Rate \& BD-PSNR. 


\bibliographystyle{IEEEtran}
\bibliography{paper}

\begin{thebibliography}{10}
\providecommand{\url}[1]{#1}
\csname url@samestyle\endcsname
\providecommand{\newblock}{\relax}
\providecommand{\bibinfo}[2]{#2}
\providecommand{\BIBentrySTDinterwordspacing}{\spaceskip=0pt\relax}
\providecommand{\BIBentryALTinterwordstretchfactor}{4}
\providecommand{\BIBentryALTinterwordspacing}{\spaceskip=\fontdimen2\font plus
\BIBentryALTinterwordstretchfactor\fontdimen3\font minus
  \fontdimen4\font\relax}
\providecommand{\BIBforeignlanguage}[2]{{%
\expandafter\ifx\csname l@#1\endcsname\relax
\typeout{** WARNING: IEEEtran.bst: No hyphenation pattern has been}%
\typeout{** loaded for the language `#1'. Using the pattern for}%
\typeout{** the default language instead.}%
\else
\language=\csname l@#1\endcsname
\fi
#2}}
\providecommand{\BIBdecl}{\relax}
\BIBdecl

\bibitem{standard:jpeg94}
{ISO/IEC 10918-1:1994}, ``{Digital Compression and Coding of Continuous-tone
  Still images: Requirements and Guidelines},'' 1994.

\bibitem{jpeg2000}
C.~Christopoulos, A.~Skodras, and T.~Ebrahimi, ``{The JPEG2000 still image
  coding system: an overview},'' \emph{IEEE Trans. on Consumer Electronics},
  vol.~46, no.~4, pp. 1103--1127, 2000.

\bibitem{sun}
D.~Sun and W.-K. Cham, ``{Postprocessing of Low Bit-Rate Block DCT Coded Images
  Based on a Fields of Experts Prior},'' \emph{IEEE Trans. on Image Proc.},
  vol.~16, no.~11, pp. 2743--2751, 2007.

\bibitem{foi}
A.~Foi, V.~Katkovnik, and K.~Egiazarian, ``{Pointwise Shape-Adaptive DCT for
  High-Quality Denoising and Deblocking of Grayscale and Color Images},''
  \emph{IEEE Trans. on Image Proc.}, vol.~16, no.~5, pp. 1395--1411, 2007.

\bibitem{bm3d}
K.~Dabov, A.~Foi, V.~Katkovnik, and K.~O. Egiazarian, ``{Image Denoising by
  Sparse 3-D Transform-Domain Collaborative Filtering},'' \emph{IEEE Trans. on
  Image Proc.}, vol.~16, pp. 2080--2095, 2007.

\bibitem{overlap}
X.~Zhang, R.~Xiong, X.~Fan, S.~Ma, and W.~Gao, ``{Compression Artifact
  Reduction by Overlapped-Block Transform Coefficient Estimation With Block
  Similarity},'' \emph{IEEE Trans. on Image Proc.}, vol.~22, no.~12, pp.
  4613--4626, 2013.

\bibitem{ren}
J.~Ren, J.~Liu, M.~Li, W.~Bai, and Z.~Guo, ``{Image Blocking Artifacts
  Reduction via Patch Clustering and Low-Rank Minimization},'' in \emph{2013
  Data Compression Conf.}\hskip 1em plus 0.5em minus 0.4em\relax IEEE, 2013,
  pp. 516--516.

\bibitem{dictv}
H.~Chang, M.~K. Ng, and T.~Zeng, ``{Reducing Artifacts in JPEG Decompression
  Via a Learned Dictionary},'' \emph{IEEE Trans. on Signal Proc.}, vol.~62,
  no.~3, pp. 718--728, 2013.

\bibitem{wnnm}
S.~Gu, L.~Zhang, W.~Zuo, and X.~Feng, ``{Weighted Nuclear Norm Minimization
  with Application to Image Denoising},'' in \emph{Proc. of the IEEE Conf. on
  Computer Vision and Pattern Recognition}, 2014, pp. 2862--2869.

\bibitem{concolor}
J.~Zhang, R.~Xiong, C.~Zhao, Y.~Zhang, S.~Ma, and W.~Gao, ``{CONCOLOR:
  Constrained Non-Convex Low-Rank Model for Image Deblocking},'' \emph{IEEE
  Trans. on Image Proc.}, vol.~25, no.~3, pp. 1246--1259, 2016.

\bibitem{dcnn}
K.~Zhang, W.~Zuo, Y.~Chen, D.~Meng, and L.~Zhang, ``{Beyond a Gaussian
  Denoiser: Residual Learning of Deep CNN for Image Denoising},'' \emph{IEEE
  Trans. on Image Proc.}, vol.~26, no.~7, pp. 3142--3155, 2017.

\bibitem{arcnn}
C.~Dong, Y.~Deng, C.~C.~Loy, and X.~Tang, ``{Compression Artifacts Reduction by
  a Deep Convolutional Network},'' in \emph{Proc. of the IEEE Int. Conf. on
  Computer Vision}, 2015, pp. 576--584.

\bibitem{framework}
F.~Jiang, W.~Tao, S.~Liu, J.~Ren, X.~Guo, and D.~Zhao, ``{An End-to-End
  Compression Framework Based on Convolutional Neural Networks},'' \emph{IEEE
  Trans. on Circuits and Systems for Video Technology}, vol.~28, no.~10, pp.
  3007--3018, 2018.

\bibitem{virtual}
L.~Zhao, H.~Bai, A.~Wang, and Y.~Zhao, ``Learning a virtual codec based on deep
  convolutional neural network to compress image,'' \emph{Journal of Visual
  Comm. and Image Rep.}, vol.~63, pp. 102\,589.1--11, 2019.

\bibitem{feng}
Y.~Li, D.~Liu, H.~Li, L.~Li, Z.~Li, and F.~Wu, ``Learning a convolutional
  neural network for image compact-resolution,'' \emph{IEEE Transactions on
  Image Processing}, vol.~28, no.~3, pp. 1092--1107, 2018.

\bibitem{fullresolution}
G.~Toderici, D.~Vincent, N.~Johnston, S.~Jin~Hwang, D.~Minnen, J.~Shor, and
  M.~Covell, ``{Full Resolution Image Compression with Recurrent Neural
  Networks},'' in \emph{Proc. of the IEEE Conf. on Computer Vision and Pattern
  Recognition}, 2017, pp. 5306--5314.

\bibitem{optimized}
J.~Ball{\'e}, V.~Laparra, and E.~P. Simoncelli, ``{End-to-end Optimized Image
  Compression},'' \emph{Int. Conf. on Learning Representations}, 2017.

\bibitem{theis}
L.~Theis, W.~Shi, A.~Cunningham, and F.~Husz{\'a}r, ``{Lossy Image Compression
  with Compressive Autoencoders},'' \emph{arXiv preprint arXiv:1703.00395},
  2017.

\bibitem{rippel}
O.~Rippel and L.~Bourdev, ``{Real-Time Adaptive Image Compression},'' in
  \emph{Proc. of the 34th Int. Conf. on Machine Learning - Volume 70}, ser.
  ICML'17.\hskip 1em plus 0.5em minus 0.4em\relax JMLR.org, 2017, p.
  2922–2930.

\bibitem{johnston}
N.~Johnston, D.~Vincent, D.~Minnen, M.~Covell, S.~Singh, T.~Chinen,
  S.~Jin~Hwang, J.~Shor, and G.~Toderici, ``{Improved Lossy Image Compression
  with Priming and Spatially Adaptive Bit Rates for Recurrent Networks},'' in
  \emph{Proc. of the IEEE Conf. on Computer Vision and Pattern Recognition},
  2018, pp. 4385--4393.

\bibitem{learning}
M.~Li, W.~Zuo, S.~Gu, D.~Zhao, and D.~Zhang, ``{Learning Convolutional Networks
  for Content-Weighted Image Compression},'' in \emph{Proc. of the IEEE Conf.
  on Computer Vision and Pattern Recognition}, 2018, pp. 3214--3223.

\bibitem{balle2}
J.~Ball{\'e}, D.~Minnen, S.~Singh, S.~J. Hwang, and N.~Johnston, ``Variational
  image compression with a scale hyperprior,'' \emph{arXiv preprint
  arXiv:1802.01436}, 2018.

\bibitem{lee}
J.~Lee, S.~Cho, and S.-K. Beack, ``{Context-adaptive Entropy Model for
  End-to-end Optimized Image Compression},'' \emph{arXiv preprint
  arXiv:1809.10452}, 2018.

\bibitem{minnen}
D.~Minnen, J.~Ball\'{e}, and G.~D. Toderici, ``{Joint Autoregressive and
  Hierarchical Priors for Learned Image Compression},'' in \emph{Advances in
  Neural Information Proc. Systems 31}, S.~Bengio, H.~Wallach, H.~Larochelle,
  K.~Grauman, N.~Cesa-Bianchi, and R.~Garnett, Eds.\hskip 1em plus 0.5em minus
  0.4em\relax Curran Associates, Inc., 2018, pp. 10\,771--10\,780.

\bibitem{mentzer}
F.~Mentzer, E.~Agustsson, M.~Tschannen, R.~Timofte, and L.~Van~Gool,
  ``{Conditional Probability Models for Deep Image Compression},'' in
  \emph{Proc. of the IEEE Conf. on Computer Vision and Pattern Recognition},
  2018, pp. 4394--4402.

\bibitem{choi}
Y.~Choi, M.~El-Khamy, and J.~Lee, ``{Variable Rate Deep Image Compression with
  a Conditional Autoencoder},'' in \emph{Proc. of the IEEE Int. Conf. on
  Computer Vision}, 2019, pp. 3146--3154.

\bibitem{yang}
F.~Yang, L.~Herranz, J.~van~de Weijer, J.~A.~I. Guiti{\'a}n, A.~M. L{\'o}pez,
  and M.~G. Mozerov, ``{Variable Rate Deep Image Compression With Modulated
  Autoencoder},'' \emph{IEEE Signal Proc. Letters}, vol.~27, pp. 331--335,
  2020.

\bibitem{chen}
T.~{Chen} and Z.~{Ma}, ``{Variable Bitrate Image Compression with Quality
  Scaling Factors},'' in \emph{ICASSP 2020 - 2020 IEEE Int. Conf. on Acoustics,
  Speech and Signal Proc. (ICASSP)}, 2020, pp. 2163--2167.

\bibitem{edsr}
B.~Lim, S.~Son, H.~Kim, S.~Nah, and K.~Mu~Lee, ``{Enhanced Deep Residual
  Networks for Single Image Super-Resolution},'' in \emph{Proc. of the IEEE
  Conf. on Computer Vision and Pattern Recognition Workshop}, 2017, pp.
  136--144.

\bibitem{vdsr}
J.~Kim, J.~Kwon~Lee, and K.~Mu~Lee, ``{Accurate Image Super-Resolution Using
  Very Deep Convolutional Networks},'' in \emph{Proc. of the IEEE Conf. on
  Computer vision and Pattern Recognition}, 2016, pp. 1646--1654.

\bibitem{edge2}
R.~K. Pandey, N.~Saha, S.~Karmakar, and A.~Ramakrishnan, ``{MSCE: An
  Edge-Preserving Robust Loss Function for Improving Super-Resolution
  Algorithms},'' in \emph{Int. Conf. on Neural Information Proc.}\hskip 1em
  plus 0.5em minus 0.4em\relax Springer, 2018, pp. 566--575.

\bibitem{edge}
G.~Seif and D.~Androutsos, ``{Edge-Based Loss Function for Single Image
  Super-Resolution},'' in \emph{2018 IEEE Int. Conf. on Acoustics, Speech and
  Signal Proc. (ICASSP)}, 2018, pp. 1468--1472.

\bibitem{kim}
B.~{Lim}, S.~{Son}, H.~{Kim}, S.~{Nah}, and K.~M. {Lee}, ``{Enhanced Deep
  Residual Networks for Single Image Super-Resolution},'' in \emph{2017 IEEE
  Conference on Computer Vision and Pattern Recognition Workshops (CVPRW)},
  2017, pp. 1132--1140.

\bibitem{imagenet}
O.~Russakovsky, J.~Deng, H.~Su, J.~Krause, S.~Satheesh, S.~Ma, Z.~Huang,
  A.~Karpathy, A.~Khosla, M.~Bernstein \emph{et~al.}, ``{Imagenet Large Scale
  Visual Recognition Challenge},'' \emph{Int. Journal of Computer Vision}, vol.
  115, no.~3, pp. 211--252, 2015.

\bibitem{live}
H.~R. Sheikh, Z.~Wang, L.~Cormack, and A.~C. Bovik, ``{LIVE Image Quality
  Assessment Database Release 2 (2005)},'' 2005.

\bibitem{set14}
R.~Zeyde, M.~Elad, and M.~Protter, ``On single image scale-up using
  sparse-representations,'' in \emph{Int. Conf. on Curves and Surfaces}.\hskip
  1em plus 0.5em minus 0.4em\relax Springer, 2010, pp. 711--730.

\bibitem{set5}
M.~Bevilacqua, A.~Roumy, C.~Guillemot, and M.~L. Alberi-Morel, ``Low-complexity
  single-image super-resolution based on nonnegative neighbor embedding,''
  \emph{BMVA press}, 2012.

\bibitem{kodak}
\BIBentryALTinterwordspacing
C.~B. MacKnight, ``{Kodak Photo CD Eastman Kodak Company Kodak Information
  Center Department E 343 State Street Rochester, NY 14650-0811},''
  \emph{Journal of Computing in Higher Education}, vol.~7, no.~1, pp. 129--131,
  Sep 1995. [Online]. Available: \url{https://doi.org/10.1007/BF02946148}
\BIBentrySTDinterwordspacing

\bibitem{clic}
\BIBentryALTinterwordspacing
{Workshop And Challenge On Learned Image Compression}, ``{CLIC 2019 dataset},''
  2019. [Online]. Available: \url{http://www.compression.cc/challenge/}
\BIBentrySTDinterwordspacing

\bibitem{classictest}
\BIBentryALTinterwordspacing
``{C}lassic 5 dataset,'' accessed December 2019. [Online]. Available:
  \url{https://github.com/cszn/DnCNN/tree/master/testsets/classic5}
\BIBentrySTDinterwordspacing

\bibitem{general100}
C.~Dong, C.~C. Loy, and X.~Tang, ``Accelerating the super-resolution
  convolutional neural network,'' in \emph{European Conf. on Computer
  Vision}.\hskip 1em plus 0.5em minus 0.4em\relax Springer, 2016, pp. 391--407.

\bibitem{adam}
D.~P. Kingma and J.~Ba, ``{Adam: A Method for Stochastic Optimization},''
  \emph{CoRR}, vol. abs/1412.6980, 2014.

\bibitem{dipti}
D.~{Mishra}, S.~K. {Singh}, and R.~K. {Singh}, ``Wavelet-based deep auto
  encoder-decoder (wdaed)-based image compression,'' \emph{IEEE Transactions on
  Circuits and Systems for Video Technology}, vol.~31, no.~4, pp. 1452--1462,
  2021.

\bibitem{dpw}
H.~Chen, X.~He, L.~Qing, S.~Xiong, and T.~Q. Nguyen, ``{DPW-SDNet}: Dual
  {Pixel-Wavelet Domain Deep CNNs for Soft Decoding of {JPEG}- Compressed
  Images},'' in \emph{Proc. of the IEEE Conf. on Computer Vision and Pattern
  Recognition Workshops}, 2018, pp. 711--720.

\bibitem{cascnn}
L.~Cavigelli, P.~Hager, and L.~Benini, ``{CAS-CNN}: {A Deep Convolutional
  Neural Network for Image Compression Artifact Suppression},'' in \emph{2017
  Int. Joint Conf. on Neural Networks (IJCNN)}.\hskip 1em plus 0.5em minus
  0.4em\relax IEEE, 2017, pp. 752--759.

\bibitem{galteri}
L.~Galteri, L.~Seidenari, M.~Bertini, and A.~Del~Bimbo, ``{Deep Generative
  Adversarial Compression Artifact Removal},'' in \emph{Proc. of the IEEE Int.
  Conf. on Computer Vision}, 2017, pp. 4826--4835.

\bibitem{psnrb}
C.~Yim and A.~C. Bovik, ``{Quality Assessment of Deblocked Images},''
  \emph{IEEE Trans. on Image Proc.}, vol.~20, no.~1, pp. 88--98, 2010.

\bibitem{bd}
G.~Bjontegaard, ``{Calculation of average PSNR differences between
  RD-curves},'' \emph{VCEG-M33}, 2001.

\bibitem{canny}
J.~{Canny}, ``{A Computational Approach to Edge Detection},'' \emph{IEEE Trans.
  on Pattern Analysis and Machine Intelligence}, vol. PAMI-8, no.~6, pp.
  679--698, 1986.

\bibitem{bsds500}
P.~{Arbeláez}, M.~{Maire}, C.~{Fowlkes}, and J.~{Malik}, ``{Contour Detection
  and Hierarchical Image Segmentation},'' \emph{IEEE Trans. on Pattern Analysis
  and Machine Intelligence}, vol.~33, no.~5, pp. 898--916, 2011.

\end{thebibliography}

\end{document}